\documentclass[a4paper,11pt]{article}
\usepackage[round,comma,authoryear]{natbib}
\usepackage[english]{babel}
\usepackage{epigraph}
\usepackage{amsmath,amsthm,amssymb}
\usepackage{graphicx}

\newtheorem{proposition}{Proposition}
\newtheorem{example}{Example}
\newtheorem{definition}{Definition}

\makeatletter
\def\old@comma{,}
\catcode`\,=13
\def,{%
  \ifmmode%
    \old@comma\discretionary{}{}{}%
  \else%
    \old@comma%
  \fi%
}
\makeatother

\title{Degrees of riskiness, falsifiability, and truthlikeness.\\ \vskip0.2em \large A neo-Popperian account applicable to probabilistic theories.}

\author{Leander Vignero \qquad Sylvia Wenmackers\\ \footnotesize leander.vignero@kuleuven.be \qquad sylvia.wenmackers@kuleuven.be \\ \footnotesize KU Leuven, Institute of Philosophy, Centre for Logic and Philosophy of Science,\\ \footnotesize Kardinaal Mercierplein 2 -- bus 3200, 3000 Leuven, \textsc{Belgium}.}

\date{July 7, 2021\\ \vskip0.2em \small (Forthcoming in \textit{Synthese}.)}

\begin{document}
\maketitle

\begin{abstract}
In this paper, we take a fresh look at three Popperian concepts: riskiness, falsifiability, and truthlikeness (or verisimilitude) of scientific hypotheses or theories. First, we make explicit the dimensions that underlie the notion of riskiness. Secondly, we examine if and how degrees of falsifiability can be defined, and how they are related to various dimensions of the concept of riskiness as well as the experimental context. Thirdly, we consider the relation of riskiness to (expected degrees of) truthlikeness. Throughout, we pay special attention to probabilistic theories and we offer a tentative, quantitative account of verisimilitude for probabilistic theories.
\end{abstract}

\epigraph{``Modern logic, as I hope is now evident, has the effect of enlarging our abstract imagination, and providing an infinite number of possible hypotheses to be applied in the analysis of any complex fact.'' --~\citet[p.~58]{Russell1914}.}

%%%%%%%%%%%%%%%%%%%%%%%%%%%%%%%%%%%%%%%%%%%%%%%%%%%%%%%%%%%%%%%%%%%
%%%%%%%%%%%%%%%%%%%%%%%%%%%%%%%%%%%%%%%%%%%%%%%%%%%%%%%%%%%%%%%%%%%
A theory is falsifiable if it allows us to deduce predictions that can be compared to evidence, which according to our background knowledge could show the prediction --- and hence the theory --- to be false.
In his defense of falsifiability as a demarcation criterion, Popper has stressed the importance of bold, risky claims that can be tested, for they allow science to make progress.
Moreover, in his work on verisimilitude, Popper emphasized that outright false theories can nevertheless be helpful on our way towards better theories.
In this article, we try to analyze the relevant notion of riskiness, and take a fresh look at both falsifiability and verisimilitude of probabilistic theories.
Our approach is inspired by the Bayesian literature as well as recent proposals for quantitative measures of verisimilitude and approximate truth (as reviewed in Section~\ref{sec:ReviewConseqApproach}).

The paper is structured as follows. In Section~\ref{sec:intro}, we set the scene by discussing an example due to \cite{Jefferys1990BayesianRandomGen}.
In Section~\ref{sec:risk}, we make explicit the dimensions that underlie the notion of riskiness. We also examine if and how degrees of falsifiability can be defined, and how they are related to various dimensions of the concept of riskiness as well as the experimental context. Furthermore, we consider a crucial difference between deterministic and indeterministic theories in terms of their degrees of falsifiability.
In Section~\ref{sec:truthlikeness}, we review consequence-based approaches to quantifying truthlikeness and, in Section~\ref{sec:TowardsAltVerisim}, we propose alternative definitions of (expected degrees of) truthlikeness and approximate truth that are better suited for probabilistic theories.
We summarize our findings in Section~\ref{sec:conclusions}.

\section{Three pundits, one true probability distribution}\label{sec:intro}
To illustrate the themes we want to explore here, we discuss an informal example given by \cite{Jefferys1990BayesianRandomGen} (in which we changed one of the numbers):
\begin{quote}
A priori, our ``surprise'' when we observe a value close to a sharp prediction is much greater than it would be if the theory made only a vague prediction. For example, consider a wholly imaginary world where stock market pundits provide standard deviations along with their predictions of market indexes. Suppose a pundit makes a prediction of the value of an index a year hence, and quotes a standard deviation of [2]\% for his prediction. We would probably be quite surprised if the actual value turned out to be within several percent of the prediction, and if this happened we might want to investigate the pundit more closely. By making a precise prediction, this pundit takes a great risk of being proven wrong (and losing our business). By the same token, when his prediction turns out even approximately correct, we are surprised, and the likelihood that we will follow his advice in the future may be increased. We would probably be less interested in a second pundit, who predicted the same value for the index as did the first, but who quoted a standard deviation of 20\%. We would doubtless have little interest at all in a third pundit who informed us only that ``the market will fluctuate,'' even though that prediction is virtually certain to be fulfilled!
\end{quote}

We could reconstruct this scenario in terms of a unique, true value of the index at the specified time. However, since we want to address probabilistic theories (which make empirically verifiable, statistical hypotheses about the world), we assume a true probability function instead. This may either be an objective chance function or an epistemic probability assessment that is well-calibrated (say, from a group of experts), both of which can be modeled in a Bayesian context. Also the subjective probabilities expressed by the pundits can be represented as such. Throughout this paper we will apply Bayesian ideas, so considering the evidence will typically lead to a revision of prior probabilities (assigned to theories or associated hypotheses) to posteriors, which can be computed via Bayes' theorem.

\begin{figure}[!htb]
\centering
  \includegraphics[width=0.9\textwidth]{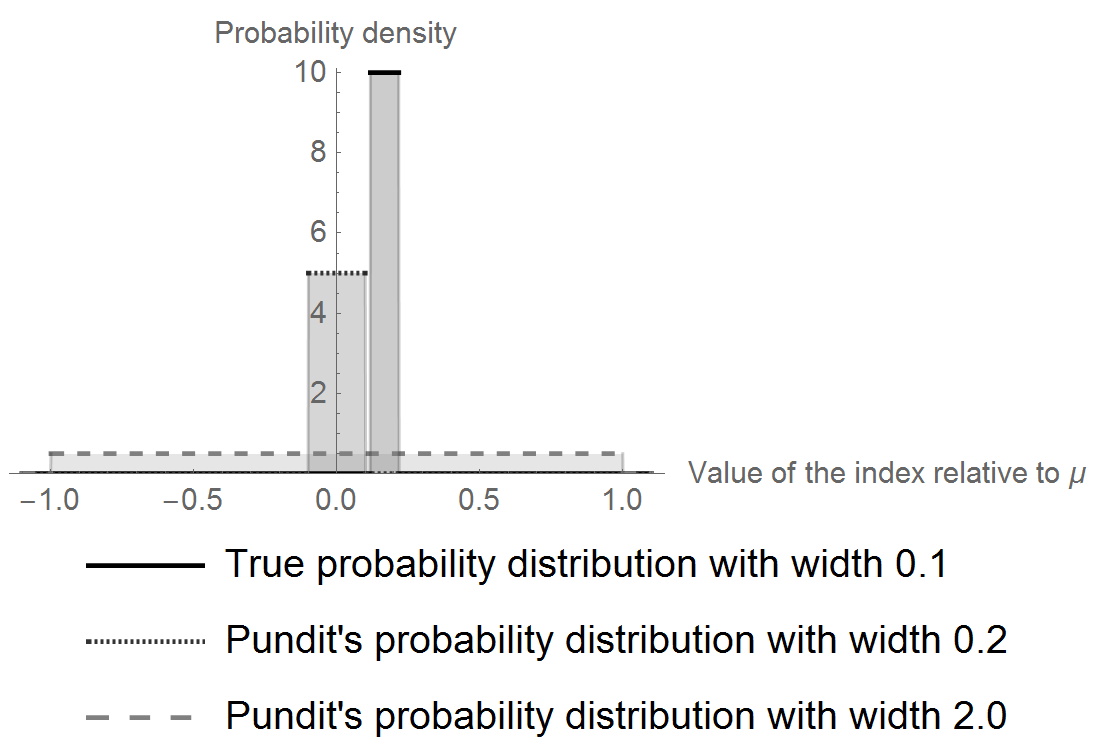}\\
  \caption{Numerical example of normalized boxcar predictions by two pundits, one more precise than the other, as well as the true distribution. The area under each of the curves is unity (due to normalization of probability distributions).}\label{Fig:PunditsStep}
\end{figure}

One modeling choice could be to use Gaussian distributions and to choose the parameters such that the first prediction is strongly disconfirmed by the evidence.
However, the results are even more striking when we opt for normalized boxcar functions (i.e., uniform distributions truncated on an interval), which allow for outright falsification of these probabilistic predictions.
So, let's assume both hypotheses of the first two pundits are normalized uniform distributions on intervals centred on the same value, $\mu$.
The first distribution is nonzero on an interval that is ten times narrower than the second.
Now assume that the interval where the true distribution takes on nonzero values and that includes the realized value turns out to fall in the second range and not in the first, but very close to the latter, as depicted in Fig.~\ref{Fig:PunditsStep}.
In this case, outright falsification of the first hypothesis occurs and the posterior probability of the second is unity (independently of the priors, as long as they start out as non-zero).
Still, we could be surprised by how close the first pundit's guess was and feel motivated to investigate it further, exactly as \cite{Jefferys1990BayesianRandomGen} described.

While the posterior for the first pundit's hypothesis is zero, it was much bolder than the second's and it was only a near miss. This may warrant looking into its neighborhood, rather than going on with the less precise, second hypothesis.
If we only consider the posteriors, however, our surprise at finding a value relatively close to the precise, first prediction seems irrational, or at least: on their own, posteriors do not offer any justification for Jefferys's move to inspect the precise, near-miss hypothesis closer.

This observation is related to another aspect that posteriors do not track: which of the competing hypotheses is more truthlike. The fact that the true hypothesis is twenty times as narrow as the second hypothesis and only half as narrow as the first or that the centre of the peak of the true hypothesis is close to that of both alternatives is not taken into account by posteriors at all. Doing this in general requires a method for measuring the similarity of the shape and position of hypothetical distributions compared to the true one.

Finally, consider the third pundit's prediction that ``the market will fluctuate'': this hypothesis is outright falsifiable (it is in principle possible that the market will turn out not to fluctuate and if this possibility is realized, this is observable), yet it is so extremely likely (on nearly any prior) that the market will fluctuate that this claim is not risky at all.
Moreover, the prediction of the third pundit has a very different structure than the former two, which could be reconstructed as given by a single probability distribution (or perhaps a narrow family thereof).
Instead, the third prediction excludes one possible realization (a constant path of the index through time), allowing all others, without assigning any probabilities to them at all.
As such, this prediction is compatible with an infinite family of probability distributions.
The negation of the third prediction corresponds to a singleton in the space of all possible paths for the index: ``the market will not fluctuate'' is equally falsifiable, but extremely risky, its prior being zero or at least extremely close to it (on nearly any probability distribution).

So, while the first two are in some sense very precise predictions, neither would be outright falsifiable or verifiable if we had chosen to reconstruct them as Gaussians (which would certainly be a defendable choice).
But the third pundit's prediction, which is a very general claim that strongly underdescribes the probabilities, is both falsifiable and verifiable.

To conclude this section, the example we borrowed from \cite{Jefferys1990BayesianRandomGen} suggests that Popper was right in emphasizing the relation between riskiness and falsifiability. Moreover, especially if we consider a Gaussian reconstruction, it seems to suggest that there are degrees of falsifiability, according to which the first pundit's claim would be falsifiable to a higher degree than the second's. The third pundit's claim, however, shows it is possible for a prediction to be outright falsifiable yet not risky at all.

%%%%%%%%%%%%%%%%%%%%%%%%%%%%%%%%%%%%%%%%%%%%%%%%%%%%%%%%%%%%%%%%%%%
\section{Riskiness and falsifiability}\label{sec:risk}
\cite{Popper1959} identified falsifiability as a demarcation criterion for scientific theories. On his account, theories that make riskier predictions are assumed to be more easily falsifiable than others.
Popper's idea of falsifiability was inspired by an asymmetry from mathematics: if a conjecture is false, it can be falsified by providing a single counterexample, whereas, if the conjecture is true, finding a proof tends to require more work. In the empirical sciences, proving a universal hypothesis (if it is indeed true) requires verifying all instances, which is impossible if the domain is infinite, whereas falsification (if the universal hypothesis is false) seems to require only a single counterexample, just like in mathematics. In practice, even for deterministic hypotheses, the picture is more complicated, due to measurement error and the Duhem--Quine problem. Moreover, for probabilistic hypotheses, falsification seems to be unobtainable (but see Section~\ref{sec:experiments}).

In any case, Popper's falsifiability, proposed as a demarcation criterion on hypotheses (or theories), is a condition on the form of the statements. Scientific hypotheses should be formulated such that empirically testable predictions can be derived from them and if the hypothesis is false, it should be possible to make an observation showing as much. This requires hypotheses to be formulated in an explicit and clear way, geared toward observational consequences. Although Popper's falsifiability was a categorical notion (either it applies to a hypothesis or it does not), not all falsifiable hypotheses are created equal: Popper preferred those that made riskier predictions.

Our first task in this section is to disentangle the various dimensions of riskiness and to track their relation to falsifiability. We also aim to formulate clear desiderata for a formal account of degrees of falsifiability, albeit without fleshing out a full proposal that achieves them.

\subsection{Two dimensions of riskiness}\label{sec:dimsriskiness}
In the context of his falsificationism, \cite{Popper1959} preferred bold or risky predictions: those that were rich in content, that were unlikely according to competing, widely accepted theories or that predicted entirely new phenomena.

On our analysis, this Popperian concept of riskiness (or boldness) consists of at least two aspects. Teasing these apart is essential for getting a good grip on the interplay between riskiness and falsifiability --- as well as between riskiness and verisimilitude (to which we turn in Section~\ref{sec:truthlikeness}). According to our analysis, the two main ingredients of riskiness are the following.
\begin{description}
\item[\textbf{(1) Informativeness}] For example, a hypothesis that gives a point prediction or a small range of possible measurement outcomes is more informative than one that gives a wide interval of possible values. That makes the former more risky. Of course, an informative prediction, which is rich in content and precise, may be wide off the mark; in other words, it may be very inaccurate, but that is precisely the point: the more informative a prediction is, the more opportunities it allows to detect discrepancies with the facts if it is false. Moreover, a substantive and precise prediction can be viewed as a conjunction of less substantive and precise hypotheses, so its prior cannot be higher than those of the latter: this is precisely what makes an informative prediction risky.
\item[\textbf{(2) Conceptual novelty}] A hypothesis may predict phenomena that have no counterpart in any previous theory and that may not have been observed yet. This aspect is related to discussions in philosophy of science concerning radically new theories, language change, etc.\ (see, e.g., \citeauthor{Mastertonetal2017}, \citeyear{Mastertonetal2017} and \citeauthor{SteeleStefanssonForthc}, \citeyear{SteeleStefanssonForthc}). Compared to the previous two aspects of riskiness, this one is more difficult to represent formally. For probabilistic theories, this issue is related to changes to the sample space, or at least the partition thereof, and it creates additional issues for how to set the priors \citep[see, e.g.,][]{WenmackersRomeijn2016}.
\end{description}

While the notion of conceptual novelty is interesting in its own right, there is still plenty of formal work to do on the first dimension, informativeness, which will be our focus here.
Informativeness is language-dependent: the granularity of the language depends on what can be expressed.
Of the three pundits in Section~\ref{sec:intro}, the first one scored the highest and the third one the lowest on this dimension. See Appendix~A for a formal development of this idea.

Informativeness does not equal improbability. To disentangle these concepts, we add a dimension of improbability that is \emph{not} a dimension of Popperian riskiness:

\begin{description}
\item[\textbf{Low probability despite equal informativeness}] Although probabilities are constrained by a partial ordering tracking informativeness, the latter is not sufficient to settle the numerical prior probability values in a unique way. Hence, subjective Bayesianism allows for individual variation across probability assignments by rational agents. These variations may be due to differences in the prior, differences in past learning or a combination thereof.
As a result, when we compare hypotheses that are equally rich in content and equally precise, they may still have unequal probabilities prior to testing according to a particular rational agent (e.g., a scientist). Advancing a hypothesis with a lower prior may be regarded as a more ``risky'' choice, but in this case, it seems irrational to pursue it. However, if we tease apart the two reasons why a hypothesis may have a low prior --- i.e., due to high informativeness and subjective variation across equally informative hypotheses --- it becomes clear that only the former source of improbability is fruitful for advancing science.
\end{description}

\subsection{Experimental context}\label{sec:experiments}
From Section~\ref{sec:dimsriskiness} we retain informativeness as an important gradable variable associated with riskiness, which makes it a promising ingredient of an analysis of falsifiability as a gradable notion. However, informativeness does not depend on the experimental context, which we analyze here.

It has become commonplace in philosophy of science and (Bayesian) epistemology to claim that probabilistic theories can only be disconfirmed to an arbitrarily high degree, but that they can never be outright falsified.
After all, the argument goes, no finite length of observations of heads falsifies the hypothesis that the coin is fair.
This example supposedly shows that there are hypotheses that are unfalsifiable, but highly disconfirmable.
However, falsification is not unobtainable for all probabilistic hypotheses in all circumstances.
In fact, we have already encountered an example of the possible falsification of a probabilistic hypothesis in Section~\ref{sec:intro}: assuming market indexes can only take on a finite number of discrete values, a distribution that is zero except for a narrow interval (such as the first pundit's curve in Fig.~\ref{Fig:PunditsStep}) is outright falsified by any observation outside of that interval.

The next two examples illustrate in yet another way why we need to fix a reference class of experiments explicitly.
They both show that even probabilistic hypotheses that do not rule out any part of the sample space in advance may still be falsifiable, given sufficient experimental resources.

\begin{example}(Emptying the bag)\label{ex:EmptyBag}
Suppose one has an opaque bag with three marbles inside: either two black marbles and one white marble or vice versa.
The only experiment allowed to gauge the color of the marbles in the bag is taking one marble out of the bag and placing it back before another draw can be made.
We might have some prior credences with regard to drawing a black marble: this constitutes a probabilistic theory.

Without the restriction to draw one marble at a time, however, there is a very simple way to find out the truth: empty the bag and see what is in it.
\end{example}

\begin{example}(Superexperiments)
Consider a demigod who can do a certain experiment an infinite number of times in a finite time frame: we call this a ``superexperiment'' --- an idea analogous to supertasks and hypercomputation.
Some theories would remain falsifiable, while others become falsifiable in this context.
Yet, a statement like ``all ravens are black'' does not become verifiable; the demigod can only test all ravens that exist at the time of the experiment, for instance.

Now, consider a jar containing a countable infinity of numbered marbles.
We know that all marbles are marked by a unique natural number; we do not know, however, whether each natural number is printed on a marble.
For instance, it could be the case that the jar only contains marbles with even numbers printed on them.
Consider the statement $h$: ``all numbers except for one unknown number, $n$, are printed on a marble in the jar.''
This hypothesis is not falsifiable by usual methods but it is falsifiable by means of a superexperiment.
Indeed, it is possible for the demigod to pick all naturals, thereby falsifying $h$.
\end{example}

The first example may strike the reader as trivial and the second as extravagant, but this is exactly the point.
These examples \citep[as well as more systematic studies along the lines of][]{Kelly1996} bring to light that we already had an implicit reference class of experiments in mind, which did not include emptying the whole bag at once or performing a superexperiment.

For a more realistic example, consider a physical theory that can only be tested by building a particle accelerator the size of the solar system.
Clearly, resource constraints make falsifying such a theory (quasi-)impossible.
This indicates that the binary distinction between possibility and impossibility of falsification does not tell the whole story and that practical constraints should be taken into account.
This general observation \citep[also made, for instance, by][]{Carroll2019} applies to probabilistic theories, too.

At this point, we hope that we have convinced the reader that the (degree of) falsifiability of a hypothesis co-depends on the severity of available experiments.
Hence, we should include the experimental context in our formal models.
Following \cite{Milne1995}, we formalize an experiment as a finite, mutually exclusive and jointly exhaustive set of propositions (equivalent to a partition of the sample space of the probability function, $P$, associated with the probabilistic hypothesis at hand). In a probabilistic context, the goal of learning from experiments is to reduce the uncertainty concerning a variable of interest; formally, this uncertainty can be modeled using an entropy measure \citep{Crupi_etal2018}. Shannon entropy is one well-known measure, but there are other options.

\subsection{Severity}\label{sec:severity}
The experimental context plays a key role in the notion of the severity of a test, for which quantifiable measures have been proposed in the literature. As discussed by \citet[pp.~60--62]{Kuipers2000}, \citeauthor{Popper1983}'s (\citeyear{Popper1983}) notion of severity refers to the ``improbability of the prediction''. Possibly, then, the qualitative notion of severity can be viewed as giving us degrees of falsifiability. This will be our working hypothesis in this section.

Severity can be defined as expected disconfirmation \citep[as proposed by][]{Milne1995} and can be related to entropy. We illustrate this in detail for an example in Appendix~B. Severity can also be related to boldness as follows: relative to a reference set of possible experiments (which could represent, say, all experiments that are possible at a given point in time), we can define the boldness of a hypothesis as the maximum severity of tests of the hypothesis. We owe this suggestion to Wayne Myrvold (personal communication), who also observed that the confirmation measure can be relativized by working through the derivation in Appendix~B from bottom to top: instead of considering the Kullback--Leibler divergence, one could start from the family of R{\'e}nyi divergences, which can be related to a family of severity measures.
On the one hand, a special case of R{\'e}nyi entropy measures is the Hartley entropy, which is of interest to our Popperian project since the expected entropy reduction associated with this measure is positive just in case the test has a possible outcome  that excludes at least one of the hypotheses under study \citep{Crupi_etal2018}. This is in line with \citeauthor{Popper1959}'s (\citeyear{Popper1959}) falsificationism, which advises learners to seek to falsify hypotheses.
On the other hand, if outright falsification is not taken to be the guiding notion, the definition of entropy could be generalized even beyond R{\'e}nyi entropies, to a two-parameter family of entropy measures \citep{Crupi_etal2018}.

Remark that, like \cite{Milne1995}, we assume that an ideal experiment induces a partition on the sample space.
So, all possible experiments can be represented by considering the lattice of partitions.
Typically, as science advances, experiments become more precise, resulting in a refinement of the experimental partition.
A more radical change in experimental context occurs when qualitatively different phenomena become measurable: this is closely related to severe theory change and, like the corresponding aspect of riskiness, we will not consider it here.

Real-world experiments are not ideal, due to measurement errors.
This, together with the Duhem--Quine problem and the issue of underdetermination of theories by empirical evidence, obviously complicates matters of outright falsification or verification.
Here we will not go into these complications, but observe that measurement error can be represented by an additional probability distribution. (For an example of a Bayesian account, see, e.g., \citeauthor{JefferysBerger1992}, \citeyear{JefferysBerger1992}.)

Our discussion so far suggests a desideratum for an adequate formal representation of the notion of gradable falsifiability.
Observe that, in the context of probabilistic theories, the language of the theory takes the shape of a partition on the sample space, as does the experimental context.
This allows for a unified treatment.
Hence, we require that degrees of falsifiability should depend on these two algebras:
\begin{enumerate}
  \item an algebra related to the language of the theory, and
  \item an algebra related to the experimental context.
\end{enumerate}

In the next section, we turn to the question of how the falsiability of probabilistic theories compares to non-probabilistic ones.

%%%%%%%%%%%%%%%%%%%%%%%%%%%%%%%%%%%%%%%%%%%%%%%%%%%%%%%%%%%%%%%%%%%
\subsection{Deterministic versus indeterministic theories}\label{sec:detvsindet}
\cite{ThyssenWenmackersForthc} proposed a classification of theories in terms of how much freedom they allow.
Although their classification was presented in a different context (the free-will debate), it is suggestive of an ordering in terms of falsifiability.\footnote{We thank Pieter Thyssen for an early discussion on this possible connection.} On their proposal, \textsc{Class I} theories are deterministic and \textsc{Class II} theories are probabilistic. However, the latter do not include all indeterministic theories. Probabilistic theories specify all possible outcomes and assign probabilities to all of them. This leaves room for two additional classes of theories: \textsc{Class III} theories allow for probability gaps and \textsc{Class IV} theories also allow for possibility gaps.
\begin{description}
  \item[\textsc{Class I}] theories are deterministic and complete. Note that a deterministic theory without free variables can be regarded as an extreme case: it assigns probability one to a single possible outcome; all other possibilities are assigned probability zero. Also note that this notion of determinism includes completeness, which is stronger than usual: incomplete theories such as ``All $F$s are $G$s'' are usually considered to be deterministic, although they remain silent on atomic possibilities; in this classification, they belong to \textsc{Class III} (see below).
  \item[\textsc{Class II}] theories are probabilistic. Within this class, some theories may assign prior probability density zero to a subset of the possibility space: the larger this set of probability zero, the closer this theory is to \textsc{Class I}. Theories that do not assign prior probability zero to any subset of the possibility space can accommodate any possible empirical observation of frequencies, albeit not all with the same probability: the more spread out the probability assignments, the lower the degree of falsifiability. All else being equal, equiprobability leads to the lowest degree of falsifiability, although it does satisfy other theoretic virtues (such as simplicity of description and symmetry).
  \item[\textsc{Class III}] theories have probability gaps: they specify all possible outcomes and may even specify relative probabilities for a subset of possible events, but they do not specify probabilities for all possible outcomes. This class includes theories that yield predictions with free variables or fudge factors that are not constrained by a probability measure. Like \textsc{Class II} theories, theories in this class can accommodate any possible empirical observation of frequencies, but they do not even assign probabilities to them. The third pundit's prediction from Section~\ref{sec:intro} (``the market will fluctuate'') belongs to this class: it does not assign probabilities to the different ways in which the market might fluctuate.
  \item[\textsc{Class IV}] theories have possibility gaps: they allow for radically new possibilities, not specified by the theory, to be realized. They may specify some possible outcomes, and even some relative probabilities of a subset of possible events, but at least under some circumstances they allow for radical openness regarding possible outcomes. The most extreme theory in this class takes the form: ``anything can happen.'' According to Popper's demarcation criterion, that is the opposite of a scientific theory, because this statement cannot be falsified by any data. Its degree of falsifiability should equal the minimal value of the scale. One could even argue that they are not theories at all, but observe that, except for the extreme case, theories belonging to this class may offer a probabilistic or even a deterministic description of certain situations.
\end{description}
This classification mainly tracks informativeness (increasing from \textsc{Class IV} to \textsc{Class I}), which we already know to correlate with riskiness and falsifiability, but it does not yet account for the experimental context.
Again, this is important: when taking into account the empirical measurement errors in a probabilistic way, even a deterministic theory will lead to a strongly spiked probability distribution at best (as already mentioned at the end of Section~\ref{sec:severity}).
That is, even though \textsc{Class I} theories are maximally falsifiable in principle, measurement errors prevent even such theories to be perfectly falsifiable in practice and, as mentioned before, the Duhem--Quine problem complicates which conclusion should be drawn even in cases where outright falsifiability seems feasible.

Let us now briefly elaborate on the possibility of empirical equivalence of theories in the light of possible outright falsification. (Alternatively, one could define this for any degree of (dis-)confirmation, but our focus in this section is on decisive evidence.)
Our approach is inspired by \citeauthor{Sklar1975}'s (\citeyear{Sklar1975}) work on ``transient underdetermination'', which refers to in-principle differences between theories that are not measurable on present empirical evidence. Likewise, we relativize the notion of empirical equivalence to a given empirical context.
In terms of falsification, two theories, $\tau_1$ and $\tau_2$, are empirically equivalent relative to an experimental context $\mathcal{E}$, if every experiment $e_i \in \mathcal{E}$ that can decide $\tau_1$ can decide $\tau_2$, and vice versa. In particular, it may happen that two theories that belong to different classes are empirically equivalent relative to the current experimental context. For example, consider two different approaches to quantum mechanics: while Bohmian mechanics belongs to \textsc{Class I}, spontaneous collapse theories belong to \textsc{Class II}. They are empirically indistinguishable relative to the current experimental context, but this need not remain the case when experimental resolution improves in the future.

Instead of only quantifying over the elements in a given class of experiments, we can also quantify over the reference classes of experiments.
For instance, if a theory $\tau_1$ can be decided by any reference class of experiments that can decide a theory $\tau_2$, we can say that $\tau_1$ is at least as decidable as $\tau_2$. If the converse holds as well, then $\tau_1$ and $\tau_2$ are empirically equivalent simpliciter (which need not imply that $\tau_1$ and $\tau_2$ are logically equivalent, of course).

At first sight, this analysis gives us no ground for preferring one among two (or more) theories that are empirically equivalent relative to the current experimental context.
However, if we take into account the possibility of future refinements (or even radical changes) to the experimental context, it may still give grounds for preferring the theory that belongs to the lowest class number (and among those, the one that is the most precise) --- provided, of course, the required future experimental improvement that distinguishes among the theories is achievable at all.
While Milne's severity measure does not take this into account, it could in principle be extended with an expectation ranging over changes of $\mathcal{E}$.

This vindicates a preference for deterministic theories over indeterministic ones, and for fully probabilistic theories over underdescribed one.
This is so, not due to any preconceptions about determinism or chance in the world, but simply because testing maximally falsifiable theories first allows us to weed out untenable options as fast as possible. Not because \textsc{Class I} theories are more likely to be true than theories of \textsc{Class II} and beyond, or even closer to the truth, but because they make maximally risky predictions and hence are maximally easy to refute when false.
This seems a solution to \citeauthor{Wheeler1956}'s problem (\citeyear{Wheeler1956}, p.~360; also quoted approvingly by \citeauthor{Popper1972}, \citeyear{Popper1972}):
\begin{quote}
  Our whole problem is to make the mistakes as fast as possible [\ldots]
\end{quote}

A similar tension pertains to the other end of the spectrum. In the face of the fallibility of science, acknowledging that there may be, as of yet, unknown possibilities seems virtuous.
This would suggest a preference for \textsc{Class IV} theories, rather than \textsc{Class I}.
However, the parts of the theory that lead to it being \textsc{Class IV} have low to minimal falsifiability.
Therefore, we argue that acknowledging the possibility of radical uncertainty should happen preferably only at the meta-theoretical level.
This prevents a conflict with the virtue of falsifiability as well as most of the difficulties in formalizing radical uncertainty. Again, this is related to the second dimension of riskiness: conceptual novelty.
Theories of \textsc{Class IV} should be considered as last resorts, if all else fails, and given the infinitude of alternatives to explore, temporary theories of this kind can always be superseded.

This concludes our discussion of degrees of falsifiability.
In the next section, we turn our attention to a different question: how to define a measure of truthlikeness?
Some of the notions that we explored in this section will reappear. For instance, it will turn out that the language of the theory is crucially important in both contexts.

%%%%%%%%%%%%%%%%%%%%%%%%%%%%%%%%%%%%%%%%%%%%%%%%%%%%%%%%%%%%%%%%%%%
\section{Formal frameworks for truthlikeness and approximate truth}\label{sec:truthlikeness}
Various authors have contributed to improving our understanding of truthlikeness and approximate truth through formal work, such as \cite{schurz1987verisimilitude,gemes2007verisimilitude,schurz2010zwart,niiniluoto2011development}, and \cite{CevolaniSchurz2017}.
As far as we know, however, these ideas have not yet been applied explicitly to probabilistic theories.

One thing we appreciate about these studies is that they show how central concepts from the traditional debates in philosophy of science can be formalized and applied to other subjects in epistemology. An illustration of this is the elegant approach to the preface paradox, presented by \cite{CevolaniSchurz2017}.

Drawing on these formal and quantitative accounts of Popperian ideas from the truthlikeness literature, we can take a fresh look at the Popperian theme from our previous section: falsifiability, which has so far eluded proper formalization and lacks a quantitative account. One of the dimensions of Popperian riskiness that we identified in Section~\ref{sec:risk} and that seems a crucial ingredient for any measure of falsifiability was informativeness. As we will see at the end of Section~\ref{sec:ReviewConseqApproach}, the content of a theory emerges as a natural measure of informativeness from the Schurz--Weingartner--Cevolani framework.

\subsection{Review of the Schurz--Weingartner--Cevolani consequence approaches}\label{sec:ReviewConseqApproach}
The aforementioned formal literature on verisimilitude has developed some quantitative ways of comparing the verisimilitude of theories as well as their approximate truth.
In particular, \cite{CevolaniSchurz2017} have proposed full definitions of verisimilitude and approximate truth, which we will restate below for ease of reference. Before we can do so, however, we have to sketch some of the background of the Schurz--Weingartner--Cevolani framework.

First, \cite{schurz1987verisimilitude} tackled verisimilitude by defining a notion of the \textit{relevant consequences of a theory}.
One can make sense of relevant consequences in both propositional languages and first-order languages.
Any given theory, $\tau$, gives rise to a set of true relevant consequences, $E_t(\tau)$, and a set of false relevant consequences, $E_f(\tau)$.
Theories are then compared by means of the sets of their relevant consequences.

Next, \cite{schurz2010zwart} built further on this account to give a quantitative definition of truthlikeness for theories represented by relevant elements (those relevant consequences that are not logically equivalent to a conjunction of shorter relevant consequences) in a propositional language with a fixed number of variables, $n$. Definition~5 in Section~5 of their paper is important for our purposes, so we include a version of it here.
This version takes into account two modifications introduced by \cite{CevolaniSchurz2017}, who added the parameter $\varphi>0$ to balance the relative weight of misses compared to matches, and normalized through division by $n$ (cf.\ Definition~9 in their paper):

\begin{definition}{Cevolani--Schurz: quantitative truthlikeness for a theory $\tau$, represented by relevant elements in a propositional language with $n$ variables.}\\
\begin{equation}\label{eq:CSverisprop}
Tr_\varphi(\tau) := \sum_{\alpha \in E_t(\tau)}Tr_\varphi(\alpha) + \sum_{\alpha \in E_f(\tau)}Tr_\varphi(\alpha),
\end{equation}
with \\
$$Tr_\varphi(\alpha) :=\left\{
\begin{array}{cl}
\frac{1}{n} \frac{v_\alpha}{k_\alpha}\frac{(n-k_{\alpha}+1)!}{n!} & \textrm{if\ } \alpha \in E_t(\tau)\\
&  \\
- \frac{\varphi}{n} \frac{(n-k_{\alpha}+1)!}{n!} & \textrm{if\ } \alpha \in E_f(\tau),%
\end{array}
\right. $$\\
where $k_\alpha$ equals the number of $\alpha$'s literals and $v_\alpha$ equals the number of $\alpha$'s true literals. The truthlikeness for Verum ($\top$) and Falsum ($\bot$) is set by the following convention: $Tr_\varphi(\top):=0$ and $Tr_\varphi(\bot):=-\varphi(n+1)/n$.
\end{definition}

On this proposal, evaluating the truthlikeness of a theory boils down to a lot of bookkeeping. The major upside of this account is its naturalness: the more true statements (of a certain kind) a theory makes, the better.
Moreover, the less contaminated these true statements are with false statements, the higher the bonus.

In addition, \cite{CevolaniSchurz2017} introduced a separate notion of approximate truth, which expresses closeness to being true, irrespective of how many other things may be true outside the scope of the hypothesis under consideration (cf.\ Definition~10 in their article).
In this context, we want to assess amounts of content, rather than bonuses versus penalties, so we suppress the subscript $\varphi$ by setting its value to unity, as follows: $Tr(\alpha) := Tr_{\varphi=1}(\alpha)$.

\begin{definition}{Cevolani--Schurz: approximate truth.}\\
 $$AT(\tau) := \frac{\sum_{\alpha \in E_t(\tau)}Tr(\alpha)}{\sum_{\alpha \in E_t(\tau)}Tr(\alpha) + \sum_{\alpha \in E_f(\tau)}|Tr(\alpha)|}.$$
\end{definition}

On this account, it is natural to think of $E_t(\tau)$ as the truth content of $\tau$ and of $E_f(\tau)$ as the falsity content of $\tau$.
The corresponding quantitative measures of truth content ($TC$) and falsity content ($FC$) of a theory $\tau$ could be defined as follows.
\begin{definition}{Quantitative content.}\\
 $$TC(\tau) := \sum_{\alpha \in E_t(\tau)}Tr(\alpha),$$
 $$FC(\tau) := \sum_{\alpha \in E_f(\tau)}Tr(\alpha),$$
 and
 $$Content(\tau) := TC(\tau) + |FC(\tau)|.$$
\end{definition}

For true $\tau_1$ and $\tau_1 \models \tau_2$, it holds that $Content(\tau_1) \geq Content(\tau_2)$, as proven in Appendix~C. This is intuitively compelling in light of the first dimension of riskiness (informativeness), which tracks the amount of content.

Using this notation, and now also suppressing $\varphi$ in the formula for quantitative truthlikeness, we can bring the essential form of the Cevolani--Schurz definitions into clear focus:
$$Tr(\tau) := TC(\tau) + FC(\tau)$$
and
$$AT(\tau) := \frac{TC(\tau)}{Content(\tau)}.$$

The measure of truth content, $TC$, is always positive and acts as a reward in the definition of truthlikeness.
The measure of falsity content, $FC$, is always negative and acts as a penalty in the definition of truthlikeness.

We can now apply these ideas to riskiness and falsifiability, as announced at the start of Section~\ref{sec:truthlikeness}. The first dimension of riskiness, informativeness, can now be understood as the $Content$ of a theory.
Ceteris paribus, as the $Content$ of a theory increases, its falsifiability increases. Notice that this provides a way of measuring content that does not depend on relative logical strength alone.
On the one hand, provided that $\tau_1$ and $\tau_2$ are true, $\tau_1 \models \tau_2$ implies $Tr_\varphi(\tau_1) \geq Tr_\varphi(\tau_2)$. (This follows from $Content(\tau_1) \geq Content(\tau_2)$, as shown above; it was also shown in \citeauthor{schurz1987verisimilitude}, \citeyear{schurz1987verisimilitude}.)
On the other hand, one can also consider theories $\tau_1$ and $\tau_2$ such that $Content(\tau_1) < Content(\tau_2)$ while $\tau_2 \not\models \tau_1$.
However, when we will turn our attention away from propositional logic towards probabilistic theories, which are a special kind of quantitative theories, merely tracking the fraction of correct assignments without measuring the distance of wrong assignments to the correct values will become implausible.

\subsection{Content of probabilistic theories}
The main thrust of the approach reviewed in the previous section is that there are meaningful notions of truth content and falsity content to begin with.
Before the advent of this approach, it seemed that content alone would never suffice to capture intuitively compelling desiderata about the structure of a truthlikeness ordering.
To see this, consider a falsehood that constitutes a near miss: for instance, a long conjunction of many true atomic conjuncts and only one false conjunct.
If we compare this to a tautology, which is the least informative kind of true statement, we would intuitively judge the former to be more truthlike than the latter.
However, \cite{tichy1974popper} and \cite{miller1974popper} have both shown that, as long as we define the content of a theory as the set of sentences closed under the consequence relation, a tautology always ranks as more truthlike than any false theory.
\cite{CevolaniSchurz2017} saves their notions of truth content and falsity content from the Tich{\'y}--Miller wrecking ball by restricting the notion of relevant consequence.
The approach to truth content and falsity content proposed by \cite{gemes2007verisimilitude} proceeds along similar lines.

Our goal here goes beyond what was already achieved by these authors: we aim to consider the truthlikeness of probabilistic theories.
An important hurdle is that there do not seem to be well-behaved counterparts to truth content and falsity content.
We now give two examples that illustrate problems at the heart of all the issues we discuss below.

\begin{example}{(Coarse-graining to a subalgebra)}\label{ex:coarse}
Assume the following probabilistic theory is true. It is characterized by the probability space $\langle \Omega, \mathcal{A}, P \rangle$ determined by the sample space $\Omega := \{A, B, C\}$, the $\sigma$-algebra $\mathcal{A} := \mathcal{P}(\Omega)$, and the probability function $P(\{A\}) := 0.2$; $P(\{B\}):=0.5$; $P(\{C\}):= 0.3$. In this probability space, the following holds: $P(\{A, B\}) = 0.7$ and $P(\{C\}) =0.3.$

There are many other probability measures over the $\sigma$-algebra $\mathcal{A}$ that satisfy this requirement, too.
This is related to the following observation: $\{A, B\}$ and $\{C\}$ uniquely determine a single subalgebra of $\mathcal{A}$: $\{\emptyset, \{A, B\}, \{C\}, \Omega \}$.

Now compare two theories probabilistic theories (which we will denote by $T$ rather than $\tau$) $T_L$ and $T_M$ that both get everything right on this subalgebra: $T_L$ says nothing else, while $T_M$ also includes probability assignments to $\{A\}$ and $\{B\}$. We are now faced with a trade-off. Clearly $T_M$ has more content, but if that content is very unlike the true theory, the truthlikeness of $T_M$ might be lower than that of $T_L$. (This can also be nicely illustrated in terms of our own proposal, as we will see in Example~\ref{ex:Linfty} in Section~\ref{sec:relativization}.)
\end{example}

Example~\ref{ex:coarse} illustrates the importance of aligning the notion of content with the ``ambient structure'', in particular, with the $\sigma$-algebra of the true probability space and coarse-grained versions of it. This is actually a quite natural requirement, in that it tracks our intuition regarding the relation between content and logical strength.
The next example shows that a probabilistic theory can be wrong on all non-trivial events and still be very close to the truth.

\begin{example}{(Wrong probabilities for all non-trivial events)}\label{example:wrong}
Building on the previous example, we can complicate matters further by considering two theories $Q$ and $R$, such that $Q(\{A\}) := 0.2; Q(\{B\}) := 0.1; Q(\{C\}):=0.7$ and $R(\{A\}) := 0.18; R(\{B\}) := 0.51; R(\{C\}) := 0.31$. Clearly, $R$ only assigns the right values to the events $\emptyset$ and $\Omega$. $Q$, on the other hand, gets the event $\{A\}$ right as well.
Yet intuitively, $R$ seems to lie closer to $P$ than $Q$ does.
\end{example}

Both examples illustrate that in tracking the truthlikeness of probabilistic theories, mere counting will no longer do. One reason is that the theories come with a native structure, which allows approximation by coarse-graining, so we cannot ignore this structure. Another reason is that the theories we are comparing are inherently quantitative, so we will have to switch from counting to measuring. This already applies for finite, discrete sample spaces --- on which we will focus here --- and is exacerbated for continuous distributions.

As a result, the orthodox consequence approach to verisimilitude no longer suffices in the probabilistic context, where verisimilitude does not depend on truth simpliciter. We need some content-specific measure to determine how close to the truth a theory really is. Notions like truth content and falsity content are therefore out of place. Issues like these just do not pop up that often when one works in propositional logic or predicate logic, so we feel that it is worthwhile to emphasize this nuance. \cite{CevolaniSchurz2017} seem to agree with us, at least tacitly, because they introduced a measure, $App$, to measure closeness in quantitative contexts, as we will see below.

\subsubsection{Representing probabilistic theories}\label{sec:app}
\cite{schurz2010zwart} are certainly right in claiming that philosophers should take knowledge representation seriously. The difference between qualitative beliefs and quantitative beliefs is so large, however, that we cannot simply transplant ideas regarding verisimilitude and approximate truth from the qualitative setting to the quantitative setting. As mentioned above, one of the key issues is that we do not have natural notions of truth content and falsity content in this setting.

The proposal of \citet[section 7.2]{CevolaniSchurz2017} is the only one we are aware of that is applicable to quantitative theories. Still, they do not consider probabilistic theories explicitly. Hence, the goal of this section is to fit probabilistic theories into the general mold they proposed for quantitative theories.

\citet{CevolaniSchurz2017} started by assuming that there is an object, $a$, that has various magnitudes, including $X_i$. The true value of this particular magnitude is assumed to be $X_i(a)=r_i^*$.
Then they suggest a quantitative theory $\tau$ is a conjunction with conjuncts of the form:
$$\tau_i : X_i(a) = r_i.$$
Furthermore, they require a measure, $App$, ranging over these conjuncts and with the following properties:
\begin{enumerate}
 \item $App(\tau_i)$ is the degree to which $r_i$ approximates the true value $r_i^*$ of the magnitude $X_i$ for object $a$.
 \item $App(\tau_i)$ ranges between $-1$ and $1$, where $App(\tau_i) = 1$ means that $r_i = r_i^*$ and $App(\tau_i) = -1$ means that the distance between $r_i$ and $r_i^*$ is maximal.
\end{enumerate}

To apply this to probabilistic theories, we consider the following choices for $a$, $X_i$ and $r_i$:
\begin{itemize}
  \item $a$ is a process or situation characterized by a probability space, including probability function $P$;
  \item $i$ indexes an event, $E_i$, from the algebra of $P$ (or an equivalent proposition);
  \item and $X_i$ is some evaluation function associated with said event, such that $X_i(a)=P(E_i)=r_i$, where $r_i$ a real number in the unit interval.
\end{itemize}

Using $App$, \cite{CevolaniSchurz2017} proposed the following measure of verisimilitude for quantitative theories:
\begin{definition}{Cevolani--Schurz: verisimilitude for quantitative theories.}\label{def:cevolani_verismilitude}\\
Let $App$ be an adequate measure that measures how well $r_i$ approximates $r_i^*$, the true value of $X_i(a)$ for the relevant $a$.
We define the verisimilitude for a conjunctive, quantitative theory $\tau$ as:
\begin{equation}\label{eq:CSverisquant}
Tr_\varphi(\tau) := \frac{1}{n}\sum_{App(\tau_i)>0} App(\tau_i) + \frac{\varphi}{n}\sum_{App(\tau_i)<0}App(\tau_i).
\end{equation}
\end{definition}
The definition in Equation~\ref{eq:CSverisquant} has the familiar form: like in Equation~\ref{eq:CSverisprop}, the first term rewards statements that approximate the truth well enough, while the second term penalizes statements that do not.

The driving force of the Schurz--Weingartner--Cevolani tradition consists of relevant consequences. Hence, we should assume that the theory $\tau$ is written down in terms of its elementary consequences.
Unfortunately, it is not exactly clear from a mathematical perspective what elementary consequences in a quantitative context comprehend. So, there remains some work to be done to determine what a probabilistic theory $\tau$ written down in terms of its elementary consequences looks like.
We assume that it includes all the probability assignments for the events in the sub-$\sigma$-algebra on which the theory is defined without the assignments of $1$ to $\Omega$ and $0$ to $\emptyset$.
\footnote{The reasoning for including all events, $E_i$, where the relevant probability measure is defined is thus:
first, if $P(E_1) = r_1; P(E_2) = r_2; P(E_1 \cap E_2) = 0$, then $P(E_1 \cup E_2) = r_1+r_2$ is a relevant consequence. Indeed, we cannot replace $E_1$ or $E_2$ by just any event \textit{salva validate}. Secondly, it does not seem the case that $P(E_1 \cup E_2) = r_1 + r_2$ is \textit{equivalent} to conjunctions of elements of the form $X_i(a) = r_i$.
For instance, $P(E_1) = r_1; P(E_2) = r_2; P(E_1 \cap E_2) = 0$ is stronger than $P(E_1 \cup E_2) = r_1 + r_2$.}\textsuperscript{,}\footnote{We have omitted the events $\emptyset$ and $\Omega$ because otherwise the trivial theory, $T_{trivial}:=(P_{trivial}(\Omega)=1; P_{trivial}(\emptyset) = 0)$, would always get assigned a high verisimilitude, which would be counterintuitive. Our choice allows us to set the default value for the trivial theories to $0$. If one is not working in a set-theoretic context, one can just replace $\emptyset$ and $\Omega$ by the $\bot$ and $\top$ elements of the algebra respectively.}

To illustrate the proposal, let us now reconsider the theories $T_L$ and $T_M$ from Example~\ref{ex:coarse}. In terms of its elementary consequences, we believe that $T_L$ should be written down as\footnote{Note that consequences of the form $L(\{A,C\}) \geq r_c$ are not elementary, since we could replace $A$ by any element of $\Omega$. We would like to thank an anonymous referee for pointing us in this direction.}
$$T_L = ( L(\{A,B\}) = r_{AB}; L(\{C\}) = r_C )$$
and that $T_M$ should be written down as
\begin{align*}
T_M = ( & M(\{A\})= r_A; M(\{B\})=r_B; M(\{C\})=r_C; M(\{ A,B \}) = r_{AB}; \\
 & M(\{ A,C \}) = r_{AC}; M(\{ B,C \}) = r_{BC} ).
\end{align*}

From now on, we will employ an abuse of notation: if $\alpha$ is a conjunct of a theory $\tau$ presented in terms of its elementary consequences, we will denote this by $\alpha \in \tau$. For instance, we can write: $M(\{ A,B \}) = r_{AB} \in T_M$.
Now let us consider $Tr_1(T_L)$ and $Tr_1(T_M)$. We can see that
\begin{equation}\label{eq:Tr1}
Tr_1(T_M) = Tr_1(T_L) + \frac{1}{n}\sum_{\tau_i \in (T_M - T_L)} App(\tau_i).
\end{equation}
In Example~\ref{ex:coarse} we assumed that $M$ was way off the mark on $A$ and $B$.
Let us assume then that $\frac{1}{n}\sum_{\tau_i \in (T_M - T_L)} App(\tau_i)$ is negative. In this case
$$Tr_1(T_M) \geq Tr_1(T_L),$$ which is what we intuitively expect. If, on the other hand, $M$ lies fairly close to $P$, then $\frac{1}{n}\sum_{\tau_i \in (T_M - T_L)} App(\tau_i)$ should be positive. In that case
$$Tr_1(T_M) \leq Tr_1(T_L),$$
which also makes sense. Let us quickly summarize: $T_L$ and $T_M$ will be assigned the same values for elementary statements that they have in common. The more fine-grained theory, $T_M$, also has the chance to get a bonus or a penalty for the statements it does not share with $T_L$.

In the preceding example, we have assumed that $App$ takes on both positive and negative values (in agreement with the second requirement for $App$ by \citeauthor{CevolaniSchurz2017}, \citeyear{CevolaniSchurz2017}). In fact, there is no straightforward way to amend the theory such that $App$ can only take negative or positive values.\footnote{\label{fn:App}Suppose that the range of $App$ would be (a subset of) $\mathbb{R}^+$. Then $Tr(T_M) \geq Tr(T_L)$, no matter how accurate or inaccurate $T_M$ is on $(T_M - T_L)$.
Indeed, on the assumption that the range of $App$ is (a subset of) $\mathbb{R}^+$, we have that $App(\tau_i) \geq 0$ for all $\tau_i \in (T_M - T_L)$. In this case, the second term of Equation~\ref{eq:Tr1}:
$$\frac{1}{n} \sum_{\tau_i \in (T_M - T_L)} App(\tau_i) \geq 0.$$
In other words, if the range of $App$ is (a subset of) $\mathbb{R}^+$, any theory $T_M$ that implies $T_L$ has a higher truthlikeness than $T_L$, irrespective of its likeness qua truth on $(T_M- T_L)$. Similarly, if the range of $App$ was (a subset of) $\mathbb{R}^-$, any theory $T_M$ that would imply $T_L$ would be lower in truthlikeness than $T_L$, irrespective of its likeness qua truth on $(T_M - T_L)$. Neither option captures how truthlikeness should behave. To avoid this, the range of $App$ should include both positive and negative real values.} Hence, we cannot simply define $App(M(E) = r_E) := |r_E - p_E|$, where $p_E$ is the value that the true probability function assigns to event $E$. Moreover, the observation also rules out defining $App$ as a statistical distance, which is a pity since it makes it harder to tie in this approach with existing work. (We return to this point in the next section.)

Again, without the full formal framework for thinking about relevant consequences in quantitative or probabilistic settings, the above account remains speculative. Further logical research is needed to assess whether the above account holds up as a concrete example of a Schurz--Weingartner--Cevolani-style theory of verisimilitude or as a variant. The ideas outlined here should make us hopeful for a well-behaved theory.

\subsubsection{Candidates for $App$}

If we try to apply Definition~\ref{def:cevolani_verismilitude} to probabilistic theories, we need to make a choice for the measure of approximation, $App$.
As shown above, the range of $App$ cannot simply be a subset of either $\mathbb{R}^+$ or $\mathbb{R}^-$, but has to take on both positive and negative values. This means that $App$ needs to distinguish elementary consequences that perform `well enough' qua likeness to truth (by assigning a positive value to them) from those that do not (by assigning a negative value).
Given that it is usually better that the proposed probability of an event $E$ lies close to the real probability of $E$, it is natural to work with a threshold: $App$ assigns positive values to probabilities that differ by less than the threshold, negative values to those that differ more, and zero to those exactly at the threshold. How large the threshold ought to be might be context-dependent.

One way to meet the criteria for such a measure is as follows, where $\epsilon \in \ ]0,1]$ is the threshold and each $T_i$ is an element of theory $T$ that assigns to event $E_i$ the following probability, $P(E_i) := r_i$:
\begin{definition}\label{def:appabs}
$$App_\epsilon(T_i) := \left\{
 \begin{array}{cl}
1 - \frac{|r_i^* -r_i|}{\epsilon} & \mbox{ if } |r_i^* -r_i| < \epsilon, \\
&  \\
- \frac{1-\frac{|r_i^* - r_i|}{\epsilon}}{1-\frac{max(r_i^*, 1-r_i^*)}{\epsilon}} & \mbox{ otherwise},%
\end{array}
\right.$$
where $r_i^*$ is the true value of $P(E_i)$.
\end{definition}

\begin{figure}[!htb]
\centering
  \includegraphics[width=0.8\textwidth]{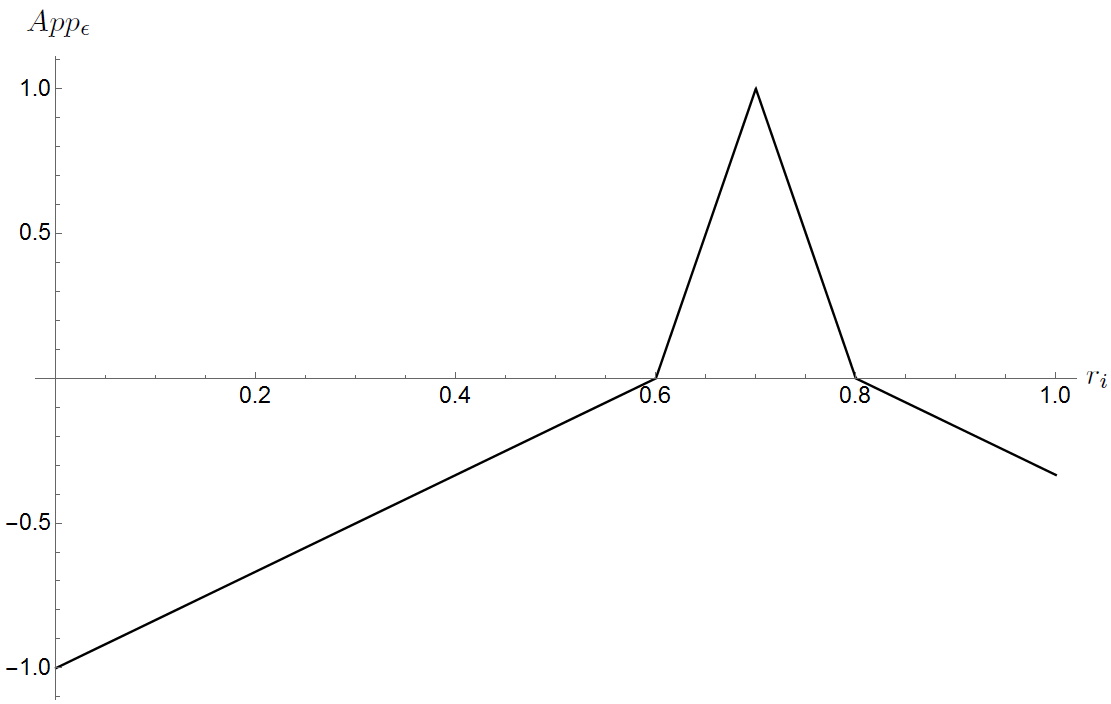}\\
  \caption{Numerical example of $App_\epsilon$ as a function of $r_i$ (the probability value that $T_i$ assigns to the event of interest), for true probability $r_i^* = 0.7$ and threshold $\epsilon = 0.1$.}\label{Fig:App}
\end{figure}

Consider, for example, an event $E_i$ with a true probability $r_i^* = 0.7$ and consider a threshold $\epsilon = 0.1$, then the corresponding graph of $App_\epsilon$ is depicted in Fig.~\ref{Fig:App}.

Alternatively, we could consider a definition in which all occurrences of $|r_i^*-r_i|$ are replaced, mutatis mutandis, by $(r_i^*-r_i)^2$ for instance.
So, while we do not find a definition for the measure of approximation that is uniquely well-motivated, at this point it might look like we have an almost complete account.

\begin{example}($App_\epsilon$-based truthlikeness for $P$ and $Q$)
Let us now illustrate this proposal for $App$ by considering the probability functions $P$, $Q$, and $R$ from Examples \ref{ex:coarse} and \ref{example:wrong}. We stipulated that $P$ was the true distribution and both $Q$ and $R$ are defined on the full algebra $\mathcal{A}$.
When we apply Definition~\ref{def:appabs} with $\epsilon = 0.1$, we obtain the numerical values of $App$ listed below:

\begin{align*}
  App(Q(\{A\})=0.2) &= 1 \hspace{40pt}  &App(R(\{A\})=0.18) &= 1 - \frac{0.02}{0.1} = 0.8 \\
   App(Q(\{B\})=0.1) &= -\frac{3}{4}    &App(R(\{B\})=0.51) &= 1 - \frac{0.01}{0.1} = 0.9 \\
    App(Q(\{C\})=0.7) &= -\frac{1}{2}    &App(R(\{C\})=0.31) &= 1 - \frac{0.01}{0.1} = 0.9 \\
    App(Q(\{A,C\} =0.9) &=-\frac{3}{4} &App(R(\{A,C\}=0.49) &= 1 - \frac{0.01}{0.1} =0.9 \\
    App(Q(\{A,B\} =0.3) &=- \frac{1}{2} &App(R(\{A,B\}=0.69) &= 1 - \frac{0.01}{0.1} =0.9\\
    App(Q(\{B,C\} =0.8) &= 1 &App(R(\{B,C\}=0.82) &= 1 - \frac{0.02}{0.1} =0.8 \\
\end{align*}

The truthlikeness for $T_R$ and $T_Q$ can now be calculated by plugging these values into Definition~\ref{def:cevolani_verismilitude}:
$$Tr(T_R) = \frac{1}{6}(0.8 + 0.9 + 0.9 + 0.9 + 0.9 + 0.8) = \frac{13}{15},$$
and
$$Tr(T_Q) = \frac{1}{6}(1 - \frac{1}{2} - \frac{3}{4} - \frac{3}{4} - \frac{1}{2} + 1) = \frac{-1}{12}.$$
We see that $Tr(T_R) > Tr(T_Q)$, as we would intuitively expect.
\end{example}

In the next section, we discuss the strengths and weaknesses of $App$-based approaches in a more general setting.

\subsection{Taking stock of consequence approaches for verisimilitude of probabilistic theories}

Initially, we were rather sceptical towards the applicability of Schurz--Weingartner--Cevolani-style approaches to verisimilitude of probabilistic theories.\footnote{We are thankful to the referees for making us reconsider our assessment.} We have since warmed up considerably to this approach and related ideas. As mentioned above, we are hopeful that further research will yield a well-behaved theory. Nevertheless, we still believe that this approach might also be plagued by some issues. These issues all
revolve around the same theme: the framework's limited compatibility with important ideas from information theory.

Firstly, $App$ forces us to compare distributions on all events that are in the set of elementary consequences. As such, $App$ as envisioned by \cite{CevolaniSchurz2017} does not seem very flexible. Indeed, if one only wants to compare distributions on a couple of specific events, one might run into problems.
Secondly, while the threshold-based construction for $App$ that we considered in the previous section is adequate (in the sense of the criteria reviewed in Section~\ref{sec:app}), it remains artificial in the context of information theory. Usually, probability distributions are compared by (pre-)distances, but there is no such ready-made candidate for $App$, because it needs to take both positive and negative values (as explained in footnote~\ref{fn:App}).

Presumably, when \cite{CevolaniSchurz2017} developed their account for quantitative theories, they had in mind measures that are qualitatively different, such as the mass and the length of an object. As such, applying it to probabilistic theories is off-label use and we should not be surprised that we obtain heterodox proposals.

The approach we develop below is not a Schurz--Weingartner--Cevolani framework. Nevertheless, we are still indebted to their approach since we compare theories in terms of their ambient logical structure. The following example will allow us to quickly revisit some of the Popperian themes.

\begin{example}{(High and low resolution)}\label{example:100vs10}
Consider a theory, $T_C$, that is defined with respect to the Boolean algebra with semantic atoms $A_1, A_2, \ldots, A_{100}$ and another theory, $T_X$, that is defined on an algebra with `atoms' $B_1, B_2, \ldots, B_{10}$, where
$$B_m=\bigcup_{i\in \{1,\ldots,10\}}A_{10(m-1)+i}.$$
($C$ and $X$ refer to the Roman numerals for 100 and 10, respectively.)

Theory $T_C$ consists of the following set of 100 statements:
$$T_C := \left\{P_C(A_i) = 0.01 \mid i \in \{1,2,\ldots,100\} \right\}.$$
Theory $T_X$ consists of the following set of 10 statements:
$$T_X := \left\{P_X(B_i) = 0.1 \mid i \in \{1,2,\ldots,10\} \right\}.$$
In this case, $T_X$ can be regarded as a low-resolution or coarse-grained version of $T_C$, since (i) the algebra of $T_X$ only contains unions of elements of the algebra of $T_C$ and (ii) the probabilities that $T_X$ assigns to these unions are equal to the sum of probabilities that $T_C$ assigns to the composing sets. In other words, where they are both defined, the probability assignments agree, and the cases where the probability is defined on $T_X$ form a strict subset of the cases where the probability is defined on $T_C$. If we assume $T_C$ is true, so is $T_X$, but $T_C$ will be assigned a higher verisimilitude than $T_X$.
\end{example}

This example is analogous to the case of non-probabilistic theories differing in information content and precision (cf.\ Section~\ref{sec:risk}).
In practice, fine-graining can come into view due to theoretical refinement or due to increases in measurement resolution.
Observe that there may also be theories that are more fine-grained in some aspects and more coarse-grained in others, so the order need not be total.

We acknowledge that the ambient logical structure and the relations it induces between theories are an essential part of a good understanding of verisimilitude and related Popperian notions. At the same time, we believe that we also need to look at the content of the theories, when considering the probabilistic case.
A probabilistic theory might have little to no (non-trivial) true consequences at all, while still being a very good approximation (recall Example~\ref{ex:coarse} and its subsequent discussion).
The crux of Popperian verisimilitude consists in notions of true and false consequence.
Indeed, \cite{schurz2010zwart} as well as \cite{Oddie2016SEP} show that Popper's account of verisimilitude can best be regarded as a consequence account. Even though we do not claim that more impure consequence accounts like \citeauthor{CevolaniSchurz2017}'s (\citeyear{CevolaniSchurz2017}) $App$-based approach are doomed to fail, we do not aim to resolve these issues in our own account either. We side-step the issue because our proposal below is not a consequence account.

%%%%%%%%%%%%%%%%%%%%%%%%%%%%%%%%%
\section{Towards an alternative definition of truthlikeness for probabilistic theories}\label{sec:TowardsAltVerisim}
At this point, we step away from the account that was based on propositional logic.
Our goal here is to come up with an adequate alternative notion of verisimilitude that takes the structure of the underlying $\sigma$-algebra (which allows coarse-graining) into account.
In Section~\ref{sec:approxtruth}, we will consider the analogous question for approximate truth.
Our main topic in this section, however, is the notion of truthlikeness in the context of probabilistic theories.
Our goal is not to give a full, formal account, which would constitute a complete research program in itself, requiring technical work akin to that of \cite{schurz1987verisimilitude} and \cite{gemes2007verisimilitude}.
Instead, our more modest aim here is twofold.
First, we want to show that we can apply some central Popperian ideas to probabilistic theories.
Secondly, we want to suggest some pitfalls and desiderata for the development of a fully fledged formal account along these lines.
We believe, in the spirit of Russell's epigraph, that propositional logic or possible-world accounts cannot hope to capture actual scientific theories; neither can the alternative we wish to prepare for.
Much like how scientists use the harmonic oscillator or other toy models and idealizations, we use formal methods to study specific aspects of science itself --- without any pretense of being able to capture all aspects of actual examples of scientific theories.
In other words, the point of our work is to expand the ``abstract imagination'' and thus to help us pinpoint fundamental issues.

\subsection{Compatibility of probabilistic theories at a certain level of grain}
We consider probabilistic theories that are fully specified by listing the probability assignments to the events that form the basis of a $\sigma$-algebra.
From here on, we will refer to probabilistic theories by their probability function.
In Example~\ref{example:100vs10}, we have encountered a case of ``compatibility'' between probability functions that were defined on algebras with a different resolution.
Let us now turn this idea into a general definition.

\begin{definition}{Compatibility of refinements.}\label{def:CompatRefine}\\
Consider two probability functions, $P$ defined on algebra $\mathcal{A}$ and $P'$ defined on algebra $\mathcal{A}'$ that is a subalgebra of $\mathcal{A}$.
We say that probability distribution $P$ is \textbf{a compatible refinement} of $P'$ if $P'(E) = P(E)$ for each event $E$ in $\mathcal{A}'$.

In this case, $P'$ is called \textbf{a compatible coarsening} of $P$; $P$ and $P'$ are said to be \textbf{compatible} with each other.
\end{definition}
So, to determine whether two probabilistic theories are compatible, we compare their probability functions at the level of the coarsest algebra among the two.
Clearly, this comparison only makes sense when the algebra of one probability function is a subalgebra of the other, as is indeed required by the definition.
Another way of coarsening that could in principle be considered is related to rounding probability values. This is not what we are dealing with here, but it could help to restrict the set of all possible probability functions to a finite set.

If we apply Definition~\ref{def:CompatRefine} to Example~\ref{example:100vs10}, we see that $P_C$ is a compatible refinement of $P_X$.
Also observe that any probability function on any non-trivial algebra is a compatible refinement of the trivial $(0,1)$-probability function on the minimal algebra (containing only the empty set and the sample space).

We denote the compatibility of a coarser probability distribution $P'$ with $P$ by $P' \leq P$, which is read as: $P'$ is a compatible coarsening of $P$.
The symbol ``$\leq$'' is fitting, since the compatible coarsening relation is reflexive ($P \leq P$ for all $P$), antisymmetric ($P' \leq P$ and $P \leq P'$ implies $P' = P$), and transitive ($P'' \leq P'$ and $P' \leq P$ implies $P'' \leq P$); hence, it is a partial order.

Given an algebra $\mathcal{A}$ and a probability function $P'$ on $\mathcal{A}'$, which is a subalgebra of $\mathcal{A}$, there may be many compatible refinements defined on an algebra $\mathcal{B}$ such that $\mathcal{A}' \subseteq \mathcal{B} \subseteq \mathcal{A}$.
We now fix a collection of compatible refinements to a given, coarser probability function in the following way.
First, we fix a finite set of probability functions on subalgebras of $\mathcal{A}$, calling it $D_\mathcal{A}$.
Then, for a probability function $P'$ defined on a subalgebra of $\mathcal{A}$, we call $D_\mathcal{A}(P')$ the subset of $D_\mathcal{A}$ consisting of distributions compatible with $P'$.

\begin{figure}[!htb]
\centering
  \includegraphics[width=0.9\textwidth]{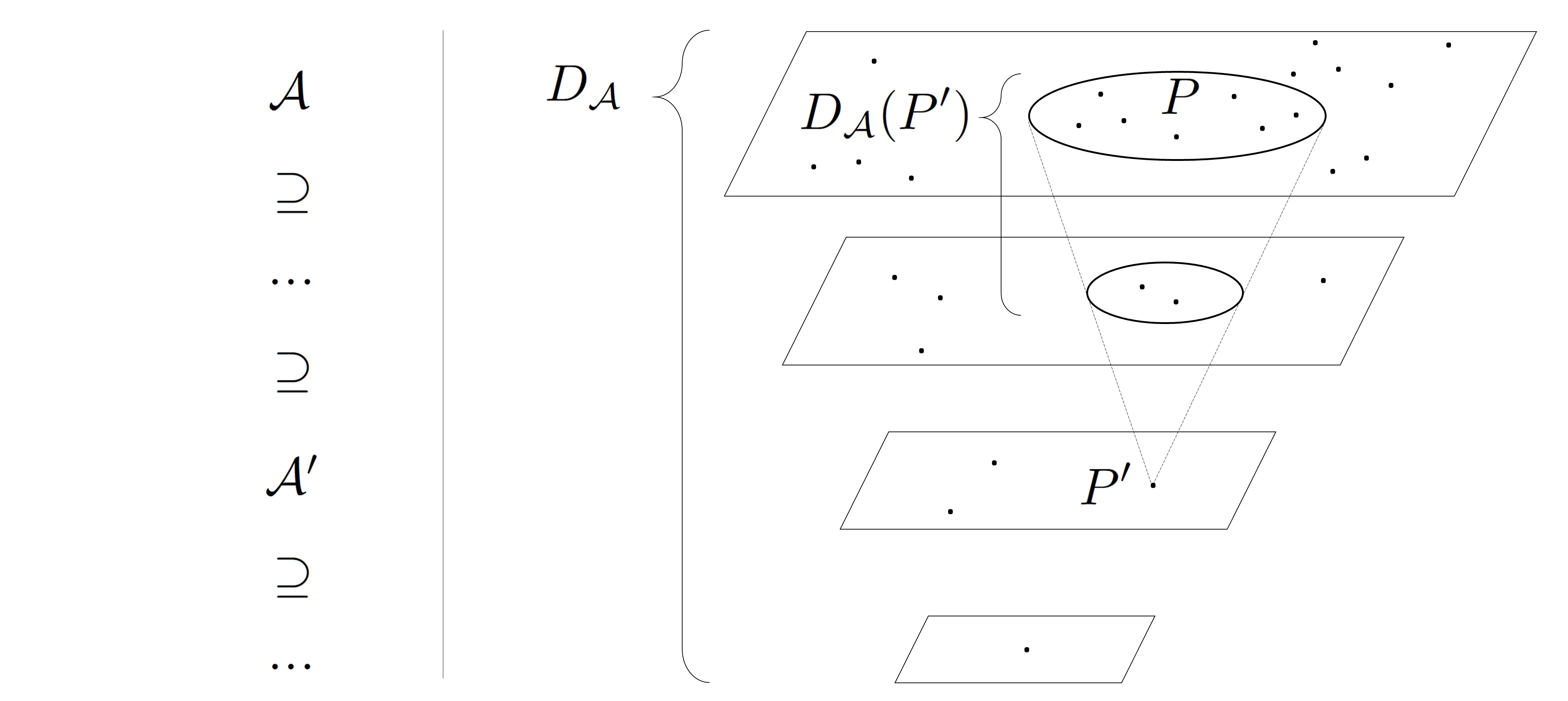}\\
  \caption{Schematic representation of a set of compatible refinements of a probability function on a coarser algebra.}\label{Fig:CompatibilitySet}
\end{figure}

Put differently,
$$D_\mathcal{A}(P') := \{P \in D_\mathcal{A} \mid P' \leq P\},$$
which is a set of compatible refinements of $P'$. The idea behind this notion is sketched in Fig.~\ref{Fig:CompatibilitySet}.

\subsection{Quantifying verisimilitude of probabilistic theories}\label{sec:quantverisim}
A natural way to define the likeness between two probabilistic theories would be to consider a statistical (pre-)distance (such as the Kullback--Leibler divergence or the Jensen--Shannon distance, to name but two examples) between their probability functions. However, these (pre-)distances are only defined if the two functions have the same domain.
We want to define the truthlikeness of a probabilistic theory, so we need to compare a hypothesized probability function with the true probability function.
Since these will typically have a different domain, we need to do a little more work first.
To achieve this, we consider the set of compatible probability functions that are defined on the same domain as the true probability function.

This way, verisimilitude can be defined as follows:

\begin{definition}{Verisimilitude of a probabilistic theory.}\label{def:versimilitude_compatible}\\
Let $P^*$ be the true probability function and $\mathcal{A}$ its algebra.
Let $D_\mathcal{A}$ be a finite set of probability functions on subalgebras of $\mathcal{A}$, with $P^* \in D_\mathcal{A}$.
Let $P'$ be a probability function on a subalgebra of $\mathcal{A}$.
Given a statistical (pre-)distance $m$, the \textbf{verisimilitude of $P'$ relative to $m$} is:\footnote{In principle, we should write $Tr_{\mathcal{A},m}$ rather than $Tr_m$, but the reference algebra $\mathcal{A}$ is usually clear from the context.}
 $$Tr_m(P') := \frac{-1}{|D_\mathcal{A}(P')|}\sum_{P \in D_\mathcal{A}(P')} m(P^*,P).$$
\end{definition}

The idea here is simple. For every $P'$, we calculate the average ``penalty'' (hence the minus sign) that the distributions in its associated set of compatible refinements $D_\mathcal{A}(P')$ accrue.
There are good reasons to opt for the average penalty here.
By proposing a theory with a more coarse-grained probability function $P'$, one casts a wider net: this way, one has a better chance to catch some of the theories that lie close to the truth ($P^*$), but one risks ensnaring some very inadequate theories as well.

The idea presented in Definition~\ref{def:versimilitude_compatible} is similar in flavor to the Tich{\'y}--Oddie average measure of verisimilitude, $Vs$ \citep{oddie1986}. Their proposal was made in a possible-world context. The verisimilitude of the theory was then calculated as unity minus the average distance from the worlds $w'$ covered by a given theory $\tau$ to the actual world $w$. In other words:
$$Vs(\tau) = 1 - \frac{1}{|\tau|}\sum_{w' \in \tau}d(w,w').$$
Instead of possible worlds, we use the probabilistic theories $P$ that are compatible with a given theory $P'$.\footnote{We would like to thank an attentive referee for pointing this out.}
Similar ideas have recently been explored by \cite{cevolani2020partial}.

Let us briefly elaborate on the fact that we have assumed the algebra $\mathcal{A}$ and the associated true probability function $P^*$ to be the most fine-grained among all probabilistic theories under consideration. This is related to an important point raised by \cite{Kuipers1982}, who distinguished between descriptive verisimilitude and theoretical verisimilitude: whereas descriptive verisimilitude tracks the closeness of a descriptive statement to the true description of the actual world, theoretical verisimilitude tracks the closeness of a candidate theory to the theoretical truth (which is compatible with more than one physical possibility). Given that we assume the true probabilistic theory (formulated in terms of $\mathcal{A}$ and $P^*$) to be maximally fine-grained, it may look as though we are dealing with the former. However, the very fact that we start from a probabilistic theory shows otherwise: by their very nature, probabilistic theories allow for multiple possibilities. To emphasize this, we can exclude theories that assign all prior probability mass to a singleton. Moreover, the structure of the algebra typically does not include the power set of the sample space. This shows that we aim to track theoretical verisimilitude rather than descriptive verisimilitude.
In addition, we allow that scientists can usefully coarse-grain relative to $\mathcal{A}$ and $P^*$, so we do not presuppose that the true probabilistic theory is at the right level of grain for all purposes.

In light of the previous section, some readers might expect us to opt for an approach analogous to introducing an $App$ function, with positive and negative contributions.
While this would in principle be possible, we have chosen otherwise. Instead, we divide by $|D_\mathcal{A}(P')|$, in order to penalize for coarseness.
This factor is also related to riskiness and falsifiability: a more coarse-grained probability function is compatible with a larger set of refinements. Hence it is less risky and has a lower degree of falsifiability.

\begin{example}($m$-based truthlikeness of $P$ and $Q$)
To illustrate Definition~\ref{def:versimilitude_compatible}, let us again consider the theories $R$ and $Q$ from Examples \ref{ex:coarse} and \ref{example:wrong}. To measure the distance between two probability distributions, $P_1$ and $P_2$ defined on the same algebra $\mathcal{A}$, we pick the Jensen--Shannon distance, $m_{JS}$.\footnote{The Jensen--Shannon distance is defined as follows:
$$m_{JS}(P_1,P_2) := \sqrt{\frac{1}{2} D_{KL}\left(P_1\ \middle\|\ \frac{1}{2} (P_1+P_2)\right)+ \frac{1}{2} D_{KL}\left(P_2\ \middle\|\ \frac{1}{2} (P_1+P_2)\right)},$$
where $D_{KL}$ refers to the Kullback--Leibler divergence, in turn defined as:
$$D_{KL}(P_1\ \|\ P_2) := \sum_{\omega \in \Omega} P_1(\{\omega\}) \log \left( \frac{P_1(\{\omega\})}{P_2(\{\omega\})} \right),$$
with the base of the logarithm an arbitrary but fixed real $>1$. We have chosen the natural logarithm for our numerical results.}
Note that in this special case, where $P$ and $Q$ are specified on the same, full algebra, the reference probability sets are singletons $D_{\mathcal{A}}(Q)=\{Q\}$ and $D_{\mathcal{A}}(R)=\{R\}$. Hence, comparing truthlikeness amounts to comparing the Jensen--Shannon distances between $P$ and $Q$, and between $P$ and $R$.
In this case, we have that
 $$Tr_{m_{JS}}(P) = \frac{-1}{1}m_{JS}(P,P) = 0,$$
 $$Tr_{m_{JS}}(R) = \frac{-1}{1}m_{JS}(P,R) \approx -0.018,$$
 $$Tr_{m_{JS}}(Q) = \frac{-1}{1}m_{JS}(P,Q) \approx -0.338.$$
Again, we find that $R$ is more truthlike than $Q$, as we would intuitively expect.
\end{example}

From the perspective of machine learning and finance, our approach looks quite familiar, especially if we consider $-Tr_m(P')$ instead of $Tr_m(P')$:
$m(P^*,P)$ plays the role of a \emph{loss function}, the average of which we want to minimize.\footnote{The philosophical community is mostly familiar with utility functions. The mirror image of these are loss functions, which are often used in finance and computer science. Maximizing utility functions corresponds to minimizing loss functions, and vice versa.}
So, $-Tr_m(P')$, which averages over the possible losses, can be considered to be a \emph{risk function} that expresses the expected utility of $P'$.

Definition~\ref{def:versimilitude_compatible} does not specify which statistical (pre-)distance $m$ one has to consider. Given that the literature on similarity between functions as well as the literature on distances between statistical objects has produced various measures, which capture different aspects of likeness, we currently do not see a uniquely well-motivated choice among them for truthlikeness either. A probabilistic theory may be very similar to the truth in some ways and less so in a different way. It may depend on the context, such as the goals of an inquiry, which of these aspects matters most. Hence, we have chosen to relativize the notion of truthlikeness to the likeness measure at hand.

Of course, there are additional variations on Definition~\ref{def:versimilitude_compatible} that we may consider.
For instance, we may increase the penalties by applying monotone functions to the (pre-)distance.
Hence, a possible alternative definition is given by:
$$Tr'_m(P') := \frac{-1}{|D_\mathcal{A}(P')|}\sum_{P \in D_\mathcal{A}(P')} \exp \left( m(P^*,P) \right).$$

\subsection{Further relativization of verisimilitude of probabilistic theories}\label{sec:relativization}
As already mentioned, more coarse-grained probability functions, $P'$, defined on a smaller subalgebra $\mathcal{A}$, cast a wider net, which we represented by $D_\mathcal{A}(P')$.
There is an alternative way to define the relevant net.
We might, for instance, be interested in all probability distributions that lie close to a given (set of) distributions.
To do this we, define the notion of $\epsilon$-compatibility.

\begin{definition}{$\epsilon$-compatibility of coarsenings.}\\
Consider an algebra $\mathcal{A}$ and a non-empty but finite set, $D_\mathcal{A}$, of probability functions on subalgebras of $\mathcal{A}$.
Consider a probability function $P \in D_\mathcal{A}$ defined on $\mathcal{A}$ and a probability function $P'$, defined on a subalgebra of $\mathcal{A}$.
Fix an $\epsilon \in \mathbb{R}^+_0$ and fix a (pre-)distance, $m$, on the space of probability distributions.
 We say that $P'$ is an \textbf{$\epsilon$-compatible coarsening} of $P$ if there exists a probability distribution $Q \in D_\mathcal{A}(P')$ such that $m(P,Q) < \epsilon$.
 In this case, we call $P$ an \textbf{$\epsilon$-compatible refinement} of $P'$.
\end{definition}

We denote the set of distributions that are $\epsilon$-compatible refinements of a given distribution $P'$ by $D_{\mathcal{A},m,\epsilon}(P')$.
This allows us to expand Definition~\ref{def:versimilitude_compatible} as follows.

\begin{definition}{$\epsilon$-verisimilitude of a probabilistic theory.}\label{def:Trmepsilon}\\
Let $P^*$ be the true probability function and $\mathcal{A}$ its algebra.
Let $D_\mathcal{A}$ be some non-empty but finite set of probability functions defined on subalgebras of $\mathcal{A}$.
Let $P'$ be a probability function on a subalgebra of $\mathcal{A}$.
 Given a statistical (pre-)distance $m$ and an $\epsilon \in \mathbb{R}_0^+$, the verisimilitude of $P'$ is:
 $$Tr_{m,\epsilon}(P') := \frac{-1}{|D_{\mathcal{A},m,\epsilon}(P')|} \sum_{P \in D_{\mathcal{A},m,\epsilon}(P')} m(P^*, P).$$
\end{definition}

This account, like that of \citet{CevolaniSchurz2017}, features a parameter. In our account, $\epsilon$ plays a very different role than $\varphi$ does in theirs: by changing $\epsilon$, we can change the scope of probability distributions that we would like to capture by our theories, as the following example shows.

\begin{example}\label{ex:Linfty}
Let us start again from Examples \ref{ex:coarse} and \ref{example:wrong} and assume that $P$, $Q$, and $R$ are distributions in some finite set of distributions that are under consideration. For ease of presentation (and especially calculation), we take the $L_\infty$ metric (which measures the distance between two discrete probability distributions as the maximum of the distance between the probability assignments over sets in the algebra). Further assume that $P'$ is defined on the algebra $\{ \emptyset, \{A,B\}, \{C\}, \Omega \}$, with $P'(\{A, B\}) := 0.7$ and $P'(\{C\}) := 0.3$. We will, again for ease of presentation, also assume that $P$, $Q$, and $R$ are the only inhabitants of $D_{\mathcal{A},L_\infty, 0.1}(P')$. We can now compare the truthlikeness of the following theories: (1) $D_{\mathcal{A},L_\infty, 0.1}(P')$, (2) $D_{\mathcal{A},L_\infty}(P')$, (3) $D_{\mathcal{A},L_\infty, 0.05}(P)$, and (4) $D_{\mathcal{A},L_\infty}(P)$.

\begin{enumerate}
\item Since $D_{\mathcal{A},L_\infty, 0.1}(P') = \{P, Q, R\}$, Definition~\ref{def:Trmepsilon} leads us to:
$$Tr_{L_\infty, 0.1}(P') = -\frac{1}{3}(0 + 0.4 + 0.02)= -0.14.$$

\item Since $D_{\mathcal{A},L_\infty}(P') = \{P, Q\}$, applying Definition~\ref{def:versimilitude_compatible} yields:
$$Tr_{L_\infty}(P') = -\frac{1}{2}(0 + 0.4) = -0.2.$$

\item Since $D_{\mathcal{A},L_\infty,0.05}(P) = \{P, R\}$, with Definition~\ref{def:Trmepsilon} we find that:
$$Tr_{L_\infty, 0.05}(P) = -\frac{1}{2}(0 + 0.02) = -0.01.$$

\item Since $D_{\mathcal{A},L_\infty}(P) = \{P\}$, Definition~\ref{def:versimilitude_compatible} shows:
$$Tr_{L_\infty}(P) = -\frac{1}{1}(0) = 0.$$
\end{enumerate}

We see all sorts of interesting behavior here. First, $Tr_{L_\infty}(P)$, the truthlikeness of the theory containing only the true theory, is the highest of all theories. This makes perfect sense: this theory has maximal likeness qua truth and maximal content. Secondly, we can compare theories that do not imply each other. For instance, we can see that $Tr_{L_\infty, 0.05}(P) > Tr_{L_\infty}(P')$. So, our proposal is not trivial in the sense that it only allows us to compare theories that stand in a certain logical relation to one another. Finally, let us have a look at how the truthlikeness of $D_{\mathcal{A},L_\infty, 0.1}(P')$ relates to the truthlikeness of $D_{\mathcal{A},L_\infty}(P')$. We can see that $Tr_{L_\infty, 0.1}(P') > Tr_{L_\infty}(P')$. This means the following: it is not necessarily the case that the truthlikeness increases as the content increases. The proposal also takes into account how close to the truth the stronger theory is. If the stronger theory has thrown out a lot of good parts of the weaker theory, it might get assigned a lower truthlikeness than the weaker theory. All of these are desirable properties for a theory of truthlikeness.
\end{example}

Moreover, Definition~\ref{def:Trmepsilon} can be extended to second-order probabilistic theories, which assign probabilities to elements of a set of (first-order) probabilistic theories, using a weighted average over $\epsilon$-verisimilitudes of the latter (using the higher-order probabilities as weight factors).

Nevertheless, there is a major downside to Definition~\ref{def:Trmepsilon}: if we coarsen $P'$ to $P''$ in such a way that $D_{\mathcal{A},m,\epsilon}(P'') - D_{\mathcal{A},m,\epsilon}(P')$ only contains distributions that have the average distance from $P^*$ to the elements of $D_{\mathcal{A},m,\epsilon}(P')$, then $Tr_m(P')=Tr_m(P'')$.
This problem occurs because averaging ``forgets'' about the size of sets. In fact, this is a problem for all approaches to verisimilitude that use averaging.

We can make verisimilitude ``remember'' the size of sets by dividing the sum of $m(P^*, P)$ by a factor smaller than $|D_{\mathcal{A},m,\epsilon}(P')|$.
In order to achieve this, we consider functions $f : \mathbb{N}_0 \to \mathbb{R}$ that satisfy the following three properties:
(1) $f(1) = 1$, (2) $f$ is increasing, and (3) $n- f(n)$ is monotonically increasing on $\mathbb{N}_0$. This entails that $f(n) \leq n$ for all $n \in \mathbb{N}_0$. We will call these functions \textit{well-behaved}.

\begin{definition}{$f$-verisimilitude of a probabilistic theory.}\label{def:versimilitude_f}\\
 Given a statistical distance $m$ and an $\epsilon \in \mathbb{R}^+_0$. Now suppose that $f$ is well-behaved. Then the $f$-truthlikeness of $P'$ is defined as follows:
 $$Tr_{m,\epsilon,f}(P') := \frac{-1}{f \left( D_{\mathcal{A},m,\epsilon}(P') \right)} \sum_{P \in D_{\mathcal{A},m,\epsilon}(P')} m(P^*, P).$$
\end{definition}

\noindent As compared to Definition~\ref{def:versimilitude_compatible}, this definition relativizes truthlikeness to two additional choices: a value for $\epsilon$ and a particular function $f$ (for example, a root function).

In fact, this ongoing relativization points us towards a crucial conceptual difficulty in quantifying versimilitude: how should one balance broadness of scope versus amount of truth?
In the case of classical logics, this problem is ``solved'' by virtue of the naturalness of the Schurz--Weingartner--Cevolani framework.
For probabilistic theories, however, there does not seem to be an equally natural choice.
In particular, it is not clear what the trade-off between coarseness of the theory and losses incurred should be.
This can be set by selecting a particular function $f$ and, arguably, this is a subjective or at least context-dependent choice.

Although Definition~\ref{def:versimilitude_f} allows for various choices, it is still not fully general, since it only applies to probabilistic theories that specify probabilities on a subalgebra of the probability function associated with the true theory.
As such, the possibility of being mistaken about the relevant sample space and the related matter of conceptual novelty have not been addressed here --- a shortcoming shared with all extant proposals for verisimilitude.

\subsection{Approximate truth of probabilistic theories}\label{sec:approxtruth}
Some of the ideas in the previous section can also be applied to the notion of approximate truth of probabilistic theories.
As before, we refer to such theories by their probability functions.
Unlike verisimilitude, approximate truth is not affected by how much or how little a theory addresses, merely which fraction of the claims that the theory does make is true.
In terms of probability functions, it does not matter how coarse the subalgebra is.
The trivial algebra (empty set and sample space) and the trivial $(0,1)$-probability assignment to it give rise to a maximal approximate truth, represented by a value of 1.
This is the same approximate truth value obtained by a theory that stipulates a probability $P'=P^*$.

As long as we are dealing with finite algebras, we can define the approximate truth value as the fraction of correct probability assignments in the basis of the function's algebra normalized by the number of elements of that basis.
However, this would again raise the problem that no distinction is made between near misses and assignments that are far off the mark.

We propose to reuse our solution from Section~\ref{sec:quantverisim}: introducing a statistical (pre-)distance between the hypothesized probability function and the compatible coarsening of the true probability function at the level of the hypothesized function. This idea is applied in the following definition.

\begin{definition}{Approximate truth of a probabilistic theory.}\label{def:ApproxTruth}\\
Let $P^*$ be the true probability function on an algebra $\mathcal{A}$ and let $P'$ be a theoretically proposed probability function defined on $\mathcal{A}' \subseteq \mathcal{A}$.
Let $D_{\mathcal{A}'}$ be a finite set of probability functions defined on $\mathcal{A}'$ (contextually defined by which probabilistic theories are relevant).
Given a statistical (pre-)distance $m$, the approximate truth of $P'$ is:
 $$AT_m(P') := 1 - \frac{m(P^*, P')}{\max \{ m(P^*, P'') \mid P'' \in D_{\mathcal{A}'} \} }.$$
\end{definition}

\subsection{Outlook}
In future work, we plan to put the proposals forwarded here to the test: by applying the proposed definitions to concrete examples, we can check how various choices of measures and parameters influence the attainment of various goals, such as convergence to the truth in the limit or speed of approach to the truth. Such a study will also help us to clarify whether the proposals in Definitions~\ref{def:versimilitude_f} and \ref{def:ApproxTruth} are robust in the following sense: different (pre-)distances $m$ usually lead to different numerical results, but subsets of them could still lead to the same ordinal ordering. The methodology for investigating this hypothesis relies on numerical simulations, similar to the work of \cite{DouvenWenmackers2017}, and requires a separate study.

Another matter that requires follow-up is more conceptual in nature: although probabilistic theories played a central role in our paper, we have not touched upon the thorny issue of the interpretation of probabilities. Yet, especially when we want to analyze the verisimilitude of probabilistic theories, it is highly relevant what the probabilities are taken to represent: are they transient features related to a lack of knowledge that may be improved upon, or are they permanent markers of irreducible stochasticity?
This question can be related to a more general one in the verisimilitude literature about the goal of science: does science aim at a complete and true description of the actual world, which can be related to a particular realization of a probabilistic model? Or does it aim at finding true laws that pick out physical possibilities among all logical possibilities?

As mentioned in Section~\ref{sec:quantverisim}, \cite{Kuipers1982} clarified that these viewpoints lead to two different notions of verisimilitude, called descriptive verisimilitude and theoretical verisimilitude, respectively.
In the context of probabilistic theories, this leads to an analogous question that can be phrased as follows: should ``the truth'' be represented as a degenerate probability distribution that assigns all probability mass to single possible outcome, which is equal to the unique physical realization that unfolds in time, or not? We have assumed that there is a most fine-grained algebra and a true probability function, $P^*$, defined on it. As such, our formalism does not presuppose that the goal of science is merely to describe the actual world, but instead to describe physical possibilities with their associated probabilities. However, if someone were to add the assumption that $P^*$ is degenerate (in the sense of being equivalent to a \textsc{Class I} theory), the same formalism may perhaps be used to make sense of descriptive verisimilitude as well.

Finally, we think these discussions can be enriched by case studies that go beyond toy problems and that consider applications outside of philosophy of science proper, for instance in computer science.

%%%%%%%%%%%%%%%%%%%%%%%%%%%%%%%%%%%%%%%%%%%%%%%%%%%%%%%%%%%%%%%%%%%
\section{Conclusions}\label{sec:conclusions}
Taking stock, in this paper we have made some progress in analyzing three Popperian concepts --- riskiness, falsifiability, and truthlikeness --- in a formal and unified context, in particular when considering probabilistic theories.
In Section~\ref{sec:risk}, we have disentangled two dimensions of riskiness. The first one, informativeness, correlates positively with gradable falsifiability and can be modeled formally. We have also clarified that a formal account of degrees of falsifiability should capture the interplay between two algebras, that of the language of the theory and that of the experimental context. We have shown that this analysis applies to deterministic and indeterministic theories, allowing for a unified treatment.
In Section~\ref{sec:truthlikeness}, we reviewed the extant proposals for a formal treatment of truthlikeness and approximate truth.
Despite the indisputable virtues of the Schurz--Weingartner--Cevolani framework, we also found some shortcomings.
One issue is that they involve a lot of bookkeeping that is, as of yet, incapable of capturing the structure of probabilistic theories.
We believe that capturing this structure is essential for measuring the ``likeness'' among theories, and for estimating the likeness to the true theory in particular.
After all, a stopped clock tells the time exactly twice in twenty-four hours, and by letting the hands turn rapidly  (even counterclockwise!) a clock can be made to indicate the right time as often as one likes. Yet, only a clock with hands that rotate clockwise at about one or twelve hours per turn can hope to be like a true time teller.
In response to some of these shortcomings, in Section~\ref{sec:TowardsAltVerisim}, we have given a general form for plausible definitions of both truthlikeness and approximate truth.

Let us now reflect on the relation between the three Popperian concepts in the title. We have seen that an important ingredient of Popperian riskiness is informativeness. Informativeness is related to both falsifiability and truthlikeness, albeit with different caveats.
In the case of falsifiability, informativeness is an important ingredient. This can be seen, for instance, from Example~\ref{example:100vs10}: there exists a lot of potentially corroborating evidence for the coarse-grained theory $T_X$ that disconfirms or falsifies the more informative theory $T_C$.
In the case of truthlikeness, riskiness increases with it provided that the theory is in fact true.
So, whereas falsifiability is unrelated to truth or falsehood of a theory, truthlikeness does depend on truth. And whereas truthlikeness is experiment-independent, falsifiability is related to experimental severity.

Hence, informativeness does not tell the whole story: the severity of available experiments should be taken into account as well.
Clearly, improbability alone is not going to cut it either: it correlates with informativeness, but also with low priors due to variation independent of logical strength and with past disconfirmation. This observation is related to an impossibility result: \cite{Sprenger2018} has shown that no corroboration measure based on statistical relevance can simultaneously represent informativeness and the severity of tests that a theory has passed successfully.
Whereas informativeness is related to improbability of priors (as far as they are correlated to logical strength), past predictive success leads to increased posterior probability: these are clearly two dimensions that cannot be meaningfully combined into one total order.

Of course, actual scientific theories and their historical development are more complicated than any formal model can hope to capture, but we think studies like ours should aim for the Russellian goal of ``enlarging our abstract imagination''.
As such, we hope our proposals will encourage further debate and development of a formal account that further unifies Popperian concepts and Bayesian or information-theoretic methods.

%%%%%%%%%%%%%%%%%%%%%%%%%%%%%%%%%%%%%%%%%%%%%%%%%%%%%%%%%%%%%%%%%%%
%%%%%%%%%%%%%%%%%%%%%%%%%%%%%%%%%%%%%%%%%%%%%%%%%%%%%%%%%%%%%%%%%%%
\section*{Acknowledgements}
We are grateful to the editors for inviting us to contribute to this topical collection, which prompted us to start this investigation.
We thank Igor Douven, Wayne Myrvold, and Wouter Termont for very helpful feedback on our paper.
We are grateful to two anonymous referees, whose detailed reports helped us to improve this paper.
LV acknowledges his doctoral fellowship of the Research Foundation Flanders (Fonds Wetenschappelijk Onderzoek, FWO) through Grant Number 1139420N; SW acknowledges funding for part of this research project through FWO Grant Number G066918N.

%%%%%%%%%%%%%%%%%%%%%%%%%%%%%%%%%%%%%%%%%%%%%%%%%%%%%%%%%%%%%%%%%%%
%%%%%%%%%%%%%%%%%%%%%%%%%%%%%%%%%%%%%%%%%%%%%%%%%%%%%%%%%%%%%%%%%%%
\section*{Appendix A: Degrees of falsifiability and logical strength}
The point of bold theories is to describe a true state of affairs with sentences of the language (say $\mathcal{L}$) that are as precise as possible.
Any observation, $\phi$, that is inconsistent with a theory falsifies it.

Here we propose a sufficient condition for relative falsifiability: theory $\tau_1$ is more falsifiable than theory $\tau_2$ if every observation that falsifies $\tau_2$ also falsifies $\tau_1$.\footnote{For the sake of simplicity, we only consider theories that have models in the first place.}
So, intuitively, theories become more falsifiable with logical strength.
Indeed, if $\tau_1$ is logically stronger than $\tau_2$, then we have the class of models $Mod(\tau_1 \cup \{\phi\}) \subseteq Mod(\tau_2 \cup \{\phi\})$ for every $\phi$.

Note that, even though we naturally prefer true theories over false ones, this does not matter for relative falsifiability: this is exactly as it should be.
Secondly, even though we believe that the above is a sufficient condition for relative falsifiability, we do not believe it to be a necessary condition. For example, \cite{Popper1963} argued that the theory of relativity is more falsifiable than psychoanalysis, even though neither seems to logically entail the other.
Our sufficient condition also plays well with our ideas regarding prior credences, since probabilities decrease with logical strength.

%%%%%%%%%%%%%%%%%%%%%%%%%%%%%%%%%%%%%%%%%%%%%%%%%%%%%%%%%%%%%%%%%%%
\section*{Appendix B: Relation between severity and entropy}
We illustrate that a severity measure corresponds to an entropy measure by means of an example. In particular, we show that the severity measure defended by \cite{Milne1995} is related to the Kullback--Leibler divergence.

\cite{Milne1995} has proposed to define the severity of a test as its expected disconfirmation.
First, suppose we have a hypothesis $h$ that we would like to test with an experiment $\mathcal{E} := \{e_1, e_2, \dots, e_N\}$.
Assume that $P(h) \neq 0$ and $P(e_i) \neq 0$ for all $e_i \in \mathcal{E}$.
Then, \cite{Milne1995} defined a measure of the degree of confirmation (first proposed by I.~J.\ Good) as follows (with the base of the logarithm arbitrary but fixed $>1$):
\begin{equation}\label{eq:cM}
c_M(h, e) := \log\left(\frac{P(h \mid e)}{P(h)}\right).
\end{equation}
Finally, he identified the severity of a test of hypothesis $h$ as the expected disconfirmation, that is:
\begin{equation}\label{eq:d}
\langle d_\mathcal{E}(h) \rangle := - \sum_{e_i \in \mathcal{E}} P(e_i) c_M(h, e_i).
\end{equation}

Milne considered $c_M(h, e)$ to be a uniquely well-chosen and natural measure of confirmation, which combines various aspects of Bayesianism and vindicates certain Popperian maxims.
Nevertheless, Milne realized the superiority claim about $c_M(h, e)$ as a measure confirmation might not convince everyone.
Indeed, in the later comparison of various measures of confirmation relative to a list of four (a-)symmetry desiderata by \cite{EellsFitelson2002}, $c_M(h, e)$ did not stand out as the best choice.
However, Milne did maintain that $c_M(h, e)$ was at least superior in another regard: as a measure of the decrease in informativeness and/or uncertainty.
This claim stood the test of time better; see, e.g., \cite{Myrvold2003}.

So, can $\langle d_\mathcal{E}(h) \rangle$, based on this measure, also be interpreted as a --- or perhaps \emph{the} --- measure of the degree of falsifiability?
Unfortunately, the controversy over $c_M(h, e)$ as an adequate measure of confirmation is not the only issue.
Observe that if there exists a possible experiment outcome, $e_i \in \mathcal{E}$, that would falsify $h$ (in the classical sense), then $P(h \mid e_i) = 0$.
In this case, however, the expected degree of confirmation accruing to $h$ with respect to experiment $\mathcal{E}$ is undefined.
Hence, the expected disconfirmation of $h$, $\langle d_\mathcal{E}(h) \rangle$, is undefined as well.
For our purposes, it is especially unfortunate that cases of outright falsification remain undefined on this approach.
As long as there is no outright falsification possible, the measure is sensitive enough to pick out differences among experiments that allow for different degrees of strong disconfirmation, but it cannot be denied that in scientific practice the possibility of outright falsification is assumed.
Even if we extend the measure to the extended reals, thus formally allowing to take on the values $\pm \infty$, this measure does not adequately distinguish between combinations of theories and experiments that are more likely to result in an outright falsification. (We discuss a qualitative classification of this kind in Section~\ref{sec:detvsindet}.)

Nevertheless, there is something intuitively compelling about Milne's proposal.
First, consider a hypothesis, $h$, and imagine an experiment, $\mathcal{E}$, that leaves the posterior upon any outcome, $e_i$, equal to the prior.
Since such an experiment does not provide any confirmation or disconfirmation, it should be assigned a minimal severity score.
We can now consider the severity of any other experiment of the same hypothesis as an assessment of how much the posterior can vary across the possible measurement outcomes as compared to the reference of such a completely insensitive experiment (and we are grateful to Wayne Myrvold for suggesting this).
Milne's proposal to use $\langle d_\mathcal{E}(h) \rangle$ is of this form, which you can see by substituting the values for an insensitive experiment (i.e., $P(h \mid e_i)=P(h)$) into the equation: this yields $\langle d_\mathcal{E}(h) \rangle = 0$, which is indeed the minimal value that this measure can attain.
While this viewpoint does not favour a particular confirmation measure, it does help to understand why, once an adequate measure of confirmation has been chosen, positing expected disconfirmation is a sensible way of assigning severity scores.

It is also interesting to consider the relation between the severity of an experiment and the boldness of a hypothesis.
A first step in the analysis is to elucidate the severity of a given experiment across two hypotheses and to relate this notion to relative entropy (in terms of the Kullback--Leibler divergence).
Consider two hypotheses, $h_1$ and $h_2$, relative to an experiment $\mathcal{E}$, and assume all relevant terms are defined.
Now suppose that the expected degree of confirmation that $h_1$ will accrue relative to $\mathcal{E}$ is higher than that of $h_2$. Or, in terms of disconfirmation:
$$\langle d_\mathcal{E}(h_1) \rangle \leq \langle d_\mathcal{E}(h_2) \rangle.$$
Applying the definitions in Equations~\ref{eq:cM} and \ref{eq:d} and Bayes' theorem yields:
$$- \sum_{i \in \{1, \ldots, N\}} P(e_i)\log\left(\frac{P(e_i \mid h_1)}{P(e_i)}\right) \leq - \sum_{i \in \{1, \ldots, N\}} P(e_i)\log\left(\frac{P(e_i \mid h_2)}{P(e_i)}\right).$$
We apply some precalculus to obtain:
$$- \sum_{i \in \{1, \ldots, N\}} P(e_i)\log\left(P(e_i \mid h_1)\right) \leq  - \sum_{i \in \{1, \ldots, N\}} P(e_i)\log\left(P(e_i \mid h_2)\right).$$
We can rewrite this in terms of the cross-entropy between two distributions, $H(\mu, \nu)$, which is only defined relative to a set of possible outcomes; here we use $\mathcal{E}$. Effectively, this requires us to restrict the algebra on which the probability functions are defined to a coarse-grained partition of the sample space. (We draw attention to this here because the interplay between various algebras and coarse-graining is a recurrent theme in our paper; see in particular the end of Section~\ref{sec:experiments} and Section~\ref{sec:TowardsAltVerisim}.)
So, assuming we restrict $P$, $P(\cdot \mid h_1)$, and $P(\cdot \mid h_2)$ to $\mathcal{E}$, we obtain:
$$H(P_\mathcal{E}, P_\mathcal{E}(\cdot \mid h_1)) \leq H(P_\mathcal{E}, P_\mathcal{E}(\cdot \mid h_2)).$$

This can be linked to elementary information theory as follows. Subtracting the Shannon entropy of $P_\mathcal{E}$, $H(P_\mathcal{E})$, from both sides of the inequality, we obtain an inequality between relative entropies (given by the Kullback--Leibler divergence, $D_{KL}$):
$$D_{KL}(P_\mathcal{E}\ \|\ P_\mathcal{E}(\cdot \mid h_1)) \leq D_{KL}(P_\mathcal{E}\ \|\ P_\mathcal{E}(\cdot \mid h_2)).$$
This means that, relative to the prior $P_\mathcal{E}$, $P_\mathcal{E}(\cdot \mid h_1)$ is more conservative than $P_\mathcal{E}(\cdot \mid h_2)$. In other words, $h_2$ is more surprising or bolder than $h_1$.
Initially, we observed that Milne's preferred measure of confirmation, $c_M$, has the drawback of being undefined for hypotheses that allow for outright falsification. It occurs when the above Kullback--Leibler divergence goes to infinity, which can indeed be regarded as a sign of bold theories.
This derivation suggests that bolder statements are easier to falsify. Or rather, they are expected to accrue less confirmation relative to a given experiment $\mathcal{E}$.

Setting aside experiments that allow outright falsification of a hypothesis for a moment, it is also interesting to observe that maximally bold hypotheses, as measured by their Kullback--Leibler divergence relative to a prior and relative to an experiment, are in a sense very precise distributions, maximally concentrated on a single atomic possibility.
This shows that boldness measured in this way (which is relative to an experiment) nicely aligns with the first dimension of riskiness: informativeness.

%%%%%%%%%%%%%%%%%%%%%%%%%%%%%%%%%%%%%%%%%%%%%%%%%%%%%%%%%%%%%%%%%%%
\section*{Appendix C: Relation between logical strength and truthlikeness}
\begin{proposition}
 Suppose that $\tau_1$ is true and that $\tau_1 \models \tau_2$. Then it holds that $Content(\tau_1) \geq Content(\tau_2)$.
\end{proposition}

\begin{proof}
 \ Conceptually, we obtain this result by noting that (1) truthlikeness should increase with logical strength and (2) the content of a true theory is just its truth content.
 Let us now do it more formally.

 Given that $\tau_1$ is true, so is $\tau_2$. This means that $\tau_1 \models E_{t}(\tau_1)$ and $E_{t}(\tau_1) \models \tau_1$.
 Furthermore, $E_f(\tau_1)$ is empty; likewise for $E_f(\tau_2)$. This means that $\tau_1 \geq_{SW} \tau_{2}$ --- for the definition of $\geq_{SW}$ consult Definition~4 of \cite{schurz2010zwart}.
 Theorem~1 from \cite{schurz2010zwart} now yields that $Tr(\tau_1) \geq Tr(\tau_2)$.
 But since $\tau_1$ and $\tau_2$ are true, $Tr(\tau_1)$ and $Tr(\tau_2)$ reduce to $Content(\tau_1)$ and $Content(\tau_2)$, respectively.
 This concludes the proof.
\end{proof}

This works nicely with one of the spearhead results of \cite{CevolaniSchurz2017}. Verisimilitude contains two parts: likeness (qua truth) and content.
If the likeness is perfect (it only pertains to true theories in the proposition above), then the only thing that can differ is the content.

%%%%%%%%%%%%%%%%%%%%%%%%%%%%%%%%%%%%%%%%%%%%%%%%%%%%%%%%%%%%%%%%%%%
%%%%%%%%%%%%%%%%%%%%%%%%%%%%%%%%%%%%%%%%%%%%%%%%%%%%%%%%%%%%%%%%%%%

\end{document}